\documentclass[12pt]{revtex4}
\usepackage{dcolumn}
\usepackage{graphicx}
\usepackage{amsmath}
\usepackage{amsfonts}
\usepackage{amssymb}
\usepackage{psfrag}
\usepackage{wrapfig}
\usepackage{subfigure}
\usepackage{makeidx}
\usepackage{bm}
\usepackage{epsf}
\usepackage{hyperref}

\begin{document}

\newcommand{\bea}{\begin{eqnarray}}
\newcommand{\eea}{\end{eqnarray}}
\newcommand{\bean}{\begin{eqnarray*}}
\newcommand{\eean}{\end{eqnarray*}}
\newcommand{\nn}{\nonumber \\}
\newcommand{\mat}[1]{\left( \matrix{#1} \right)}
\newcommand{\tmat}[1]{{\scriptsize \mat{#1}}}
\newtheorem{theorem}{\sf THEOREM}
\def\thetheorem{\thesection.\arabic{theorem}}

\def\IZ{\mathbb{Z}}
\def\IR{\mathbb{R}}
\def\IQ{\mathbb{Q}}
\def\IC{\mathbb{C}}
\def\IP{\mathbb{P}}
\def\O #1{\overline{#1}}
\def\D #1{\dot{#1}}
\def\W #1{\widetilde{#1}}
\def\WH #1{\widehat{#1}}

\def\rightaction#1{\stackrel{\rightarrow}{\partial \over \partial #1}}
\def\leftaction#1{\stackrel{\lefttarrow}{\partial \over \partial #1}}
\def\func#1{\mathop{\rm #1}\nolimits}
\def\abs#1{\left| #1\right|}
\def\braket#1{\left\langle #1 \right\rangle}
\def\bra#1{\left\langle #1\right|}
\def\ket#1{\left| #1\right\rangle}
\def\gb #1{ \left\langle #1 \right]}
\def\tgb #1{ \left[ #1 \right\rangle}

\def\bbar#1{ \overline #1}
\def\Tr{\mathop{\rm Tr}}
\def\det{\mathop{\rm det}}
\newcommand{\fref}[1]{Figure~\ref{#1}}
\def\eref#1{(\ref{#1})}
\def\bit#1{\begin{array}{c} \\ \\ \end{array} \hspace{#1 cm}}
\def\d{{\rm d}}
\def\wt{\widetilde}
\def\wtl{\widetilde{\lambda}}
\def\wh{\widehat}
\def\wht{\widehat{x}^0}
\def\whx{\widehat{x}^1}
\def\th{{\theta}}
\def\bth{{\overline{\theta}}}
\def\a{{\alpha}}
\def\ba{{\overline{\alpha}}}
\def\da{{\dot{\alpha}}}
\def\b{{\beta}}
\def\db{{\dot{\beta}}}
\def\c{{\gamma}}
\def\dc{{\dot{\gamma}}}

\def\d{\partial}
\def\rmd{{\rm d}}
\def\la{\lambda}
\def\eps{\epsilon}
\def\lblb{({\bar\la}{\bar\la})}
\def\ald{{\dot\alpha}}
\def\bed{{\dot\beta}}
\def\gad{{\dot\gamma}}
\def\sid{{\dot\rho}}
\def\vev{\braket}
\def\tgb #1{ \left[ #1 \right\rangle}
\def\bket#1{\left| #1\right]}
\def\bbra#1{\left[ #1\right|}
\def\bvev#1{\left[ #1 \right]}
\def\Label#1{\label{#1}%

\smash{\hbox to0pt{\raise1ex\hbox{\tiny[#1]}\hss}}}
\newcommand{\bbibitem}[1]{\bibitem{#1}\marginpar{#1}}

\def\Spaa{\vev}
\def\Spbb{\bvev}
\def\Spab{\gb}
\def\Spba{\tgb}

\title{General tree-level amplitudes by factorization limits}

\author{ Kang Zhou, Chenkai Qiao\footnote{The
unconventional order of authors is merely to satisfy the outdated requirement
for Phy. Degree of the school.}}

\address{Zhejiang Institute of Modern Physics, Zhejiang
University, Hangzhou, 310027, China}

\begin{abstract}
To find boundary contributions is a rather difficult
problem when applying the BCFW recursion relation. In this paper, we
propose an approach to bypass this problem by calculating general
tree amplitudes that contain no polynomial using factorization limits. More explicitly,
we construct an expression iteratively, which produces correct
factorization limits for all physical poles, and does not contain other poles, then it should be
the correct amplitude. To some extent, this approach can be considered
as an alternative way to find boundary contributions. To demonstrate
our approach, we present several examples: $\phi^4$
theory, pure gauge theory, Einstein-Maxwell theory, and Yukawa theory.
While the amplitude allows the existence of polynomials
which satisfy correct mass dimension and helicities,
this approach is not applicable to determine the full amplitude.

\end{abstract}


\maketitle
\footnotesize

\section{Introduction}

The importance of scattering amplitudes can never be overestimated
in high-energy physics,  for it serves as the intermediary
between theories and experiments. The traditional approach for the
analytic calculation of scattering amplitudes relies on Feynman
diagrams and Feynman rules, which is well systemized and has
clear physical pictures. However, with increasing number of external
states, the fast growth in the number of diagrams makes the
computation extremely complicated. Naturally, more efficient
approaches are desired.

Initiated by Witten's twistor string program \cite{Witten:2003nn},
many powerful approaches have been developed in the past decade. Among
these, the BCFW on-shell recursion  relation \cite{Britto:2004ap,
Britto:2005fq} has been successfully applied in many contexts
involving massless particles at tree and loop levels, as well as for
massive particles at tree level (see reviews  \cite{Bern:2007dw,
Feng:2011np, Elvang:2013cua} and relevant citations). In the
derivation of the recursion relation, one deforms a
pair of external momenta$^{[1]}$\footnotetext[1]{There are
other deformations, see  \cite{Risager:2005vk, Benincasa:2007qj}.} in terms of a single complex variable $z$, thus the
on-shell amplitude $A(z)$ becomes a rational function of $z$. The
behavior of $A(z)$ in the limit $z\to \infty$ becomes crucial. If
$A(z)\to 0$ when $z\to \infty$, amplitudes can be reconstructed  by
summing over residues of poles at finite positions. However, if $A(z)$
does not vanish at infinity, the boundary contribution will emerge.
Thus to get correct amplitudes, we need  to find these boundary
contributions.

It has been clarified that for many theories, such as gauge
theory  and gravity, boundary terms can be zero
with some proper choices of momentum
deformations \cite{ArkaniHamed:2008yf, Cheung:2008dn}. However, there
are other theories in which boundary contributions cannot be
avoided, for example, $\phi^4$ theory and theories with Yukawa
couplings \cite{ArkaniHamed:2008yf}. Several attempts have been proposed
for finding boundary contributions. The first one is to add auxiliary
fields so that boundary terms for the enlarged theory are
zero \cite{Benincasa:2007xk, Boels:2010mj}. By proper reduction one
gets the desired amplitudes. The second one is to analyze Feynman diagrams
carefully to isolate boundary contributions within these
diagrams \cite{Feng:2009ei, Feng:2010ku, Feng:2011twa}. With this
information, boundary terms can be calculated directly or
recursively. The third one is to relate boundary terms to zeros of
amplitudes, \textit{i.e.}, roots of amplitudes \cite{Benincasa:2011kn,
Benincasa:2011pg, Feng:2011jxa}. However, it is not easy to find
such zeros. Despite of progress mentioned above, a general effective
approach to handle boundary terms is still lacking.

In this paper, we propose an approach to calculate tree
amplitudes without polynomials,  which avoids the direct computation of  boundary
contributions. The idea is to seek an expression that is
consistent with factorization limits for all physical poles and does not contain other poles. The searching can be done iteratively.
We will start
with a scalar function which has correct factorization limits for
some poles. This starting
function can be obtained by calculating the factorization limit for
one channel, or be chosen as the result given by the BCFW recursion
relation regardless of the existence of boundary contributions.
Having this input, at each step we consider the factorization limit for a new
channel, and adjust the starting function to include it, without disturbing correct factorization limits that have been already satisfied.
When correct factorization limits for all physical channels are
included, we should eliminate possible spurious poles. Then we claim that the correct amplitude is found . This
approach disregards boundary contributions, therefore it can be
applied to circumstances in which the BCFW approach is difficult.
This approach focuses on the pole structures of amplitudes therefore it cannot detect polynomial terms which do not have any pole.
Thus, if the amplitude admits polynomials which satisfy correct mass dimension and helicities,
this ambiguity will arise and the full amplitudes cannot be determined.

This paper is organized as follows. In section 2 we give a brief
overview of this approach. Then we use it to calculate
amplitudes of $\phi^4$ theory, pure gauge theory, and the
Einstein-Maxwell theory, as shown in section 3, section
4 and section 5 respectively. In these examples, we will not encounter any spurious pole when obtain an expression provides correct factorization limits for all physical poles. In section 6, we take the four point amplitude of Yukawa theory as an example to discuss how to remove possible spurious poles. In section 7, a brief conclusion
is given.

Throughout this paper, we use the
QCD  conventions, \textit{i.e.}, $2k_i\cdot
k_j\equiv\Spaa{i|j}\Spbb{j|i}$, and
$s_{ij\cdots l}$ denotes $(k_i+k_j+\cdots+k_l)^2$.
Also $A_{s_1\cdots s_n}$ denotes an
expression which has correct factorization limits for poles $s_1,\cdots,s_n$.
Furthermore, we will neglect the
overall factor $i$ in amplitudes, consequently the corresponding
factorization is $A\to -A_L{1\over P^2}A_R$, rather than $A\to
A_L{i\over P^2}A_R$. It also implies we should take
\bea
\mathrm{Res}\Big({A(z)\over z}\Big)\Big|_{z=z_\alpha}=-\sum_h A_L(z_\alpha){1\over P^2}A_R(z_\alpha),
\eea
when using the BCFW approach.

\section{Outline of the approach \label{2}}

In this section, we present a brief discussion about the approach
used in this paper. It bases on the property
that a correct amplitude has consistent factorization limits for all
physical poles. Since a meromorphic function is uniquely determined
by its poles and related residues, if an expression has correct
factorization limits for all physical poles, and does not contain other poles, the expression is almost the correct
amplitude without polynomials that we are seeking for. Under this observation,
one can reconstruct amplitudes by imposing  consistent factorization
limits for all physical channels.

To find (or guess) the correct expression, we can start from a scalar
function depending on external momenta and helicities, which gives correct
factorization limits for some channels. Such a function can be
obtained by direct computation of the factorization limit for one
channel. For example, consider the channel $\Spaa{1|2}\to 0$ , we can
write the initial function as$^{[2]}$\footnotetext[2]{It is not necessary to sum over helicities of the
on-shell internal line, since to get  non-zero three-point
sub-amplitudes under the limit $\Spaa{1|2}\to 0$, only one kind of
helicity configurations is allowed.}
\bea
A_{\Spaa{1|2}}=-{\lim_{\Spaa{1|2}\to 0}A_L A_R\over s_{12}}=
{\lim_{\Spaa{1|2}\to 0}s_{12}A\over s_{12}}.~~\label{defin}
\eea
Obviously, it has the correct factorization limit for $\Spaa{1|2}\to
0$. The initial function can also
be chosen as the result by the BCFW approach regardless of the existence of the boundary contribution.
In the former choice, the
function provides the correct factorization limit for the
corresponding channel. In the latter choice, the function at least
provides correct factorization limits for poles detected  by the BCFW
deformation.

At this stage, we need to point out a subtlety of this algorithm.
There are many different expressions which are equivalent to
each other under some particular factorization limits. For example, under the
limit $\Spaa{1|2}\to 0$,  ${\Spaa{1|3}\over
\Spaa{1|4}}={\Spaa{2|3}\over \Spaa{2|4}} $, but without imposing the
limit, ${\Spaa{1|3}\over \Spaa{1|4}}$ and
${\Spaa{2|3}\over \Spaa{2|4}} $ are different. More generally, we
will have $f\sim f+\Spaa{1|2} g$ for arbitrary functions $f$ and $g$. Thus
when we use our algorithm, we need to choose a representative
element at each step from the entire equivalent class (category)
under some factorization limits.

Having the starting expression, the next step is to consider the
factorization limit for a new channel. For instance, we start with
\eref{defin}, and consider another channel, for example
$\Spaa{1|3}\to 0$. If
\bea \lim_{\Spaa{1|3}\to 0}s_{13}A_{\Spaa{1|2}}= \lim_{\Spaa{1|3}\to
0}A_LA_R, \eea
\textit{i.e.}, $A_{\Spaa{1|2}}$ also gives the correct factorization limit for the
pole $\Spaa{1|3}$, we move on to include the correct factorization limit for another new physical pole. If this fails, we then construct
\bea
A'_{\Spaa{1|3}}=A_{\Spaa{1|2}}+\Big(A_{\Spaa{1|3}}-{\lim_{\Spaa{1|3}\to
0}s_{13} A_{\Spaa{1|2}}\over s_{13}}\Big),~~~\label{frame-A13}\eea
where
\bea
A_{\Spaa{1|3}}=-{\lim_{\Spaa{1|3}\to 0}A_L A_R\over s_{13}}.
\eea
Now we need to see if $A'_{\Spaa{1|3}}$ has the correct factorization
limit for $\Spaa{1|2}\to 0$. If it does, we are content and move on to
a new pole. If it does not, it means the original expressions
$A_{\Spaa{1|2}}$ or $A_{\Spaa{1|3}}$, or both are not proper
choices. We need to deform them properly, \textit{i.e.}, to adopt different
representations as discussed in the previous paragraph. The goal is
that while it gives the correct factorization limit for the new pole, it
also keeps correct factorization limits for poles in earlier steps.
Although we do not have a general guidance for the choice of proper
expressions, in the following sections, we will use many examples to
demonstrate how to make efficient choices. However, choices in these examples depend on specific theories, it is not yet clear that whether analogous choices can be applied
universally to any theory.$^{[3]}$\footnotetext[3]{Although taking efficient choices will simplify the calculation, one can try to achieve this goal
`blindly' by using the following observation: The uncertainty is due to the rational function of mass dimension zero, which is helicity neutral for all
external particles and reduces to 1 in the factorization limit. Thus one can construct basis of such rational functions and fix their coefficients using other factorization limits. This is only a tentative suggestion, which is beyond the scope of this manuscript and we will leave it to the future work.}

Iterating the procedure above, we can include at least one new pole at each step. Since
with proper choices of representative expressions, the set of poles
that have correct factorization limits is enlarged, within finite steps, we
will obtain a result that has correct factorization limits for all
physical poles.

It is possible that the obtained expression also contains spurious poles. One approach to eliminate them is use our framework \eref{frame-A13} again.
Let us assume that in \eref{frame-A13}, $A_{\Spaa{1|2}}$ has correct factorization limits
for all physical poles and contains a spurious pole $\Spaa{1|3}$.
In such case, we have $A_{\Spaa{1|3}}=0$. To continue, notice that
in \eref{frame-A13}, not only $A_{\Spaa{1|2}}$ and $A_{\Spaa{1|3}}$ can be deformed, but also the expression
of $\lim_{\Spaa{1|3}\to
0}s_{13} A_{\Spaa{1|2}}$ is not unique. Thus we can deform $\lim_{\Spaa{1|3}\to
0}s_{13} A_{\Spaa{1|2}}$ under the limit $\Spaa{1|3}\to
0$, so that ${\lim_{\Spaa{1|3}\to
0}s_{13} A_{\Spaa{1|2}}\over s_{13}}$ does not contain any physical pole. Then we get the result $A'_{\Spaa{1|2}}$ which has correct factorization limits for all physical poles, while the spurious pole $\Spaa{1|3}$ has been excluded. Again, although we will demonstrate this technique in examples, there is no general guidance of how to deform $\lim_{\Spaa{1|3}\to
0}s_{13} A_{\Spaa{1|2}}$ correctly. Iterating this procedure to remove all spurious poles, we find the full amplitude as desired.

It is worth noticing that, this approach is based on the assumption that the amplitude does not contain any polynomial. If an amplitude contains a polynomial that has no pole, for instance a constant, this term cannot be detected by any factorization limit. An example is adding a $\phi^6$ term in the original $\phi^4$ Lagrangian, the $\phi^6$ term adds a constant term into the six-point amplitude of pure scalars, then such a amplitude cannot be fully calculated by our approach. However, all examples computed in following sections do not contain any polynomial term. We will give a brief proof for the absence of polynomial terms in appendix A.

The calculation of this approach is more complicated than the BCFW one
since all  possible factorization channels need to be considered,
and expressions of factorization limits also need to be fixed. However,
since factorization is a general property of amplitudes, this
approach can be applied to any quantum field theories.

\section{Example 1: $\phi^4$ theory \label{3}}

Given the general framework in the previous section, let us consider a
simplest example, the color ordered  massless $\lambda \phi^4$
theory. In this theory, the lowest-point amplitude is given by
\bea
A_4(1,2,3,4)=-\lambda.
\eea
From now on we will drop out the coupling constant $-\la$.  We will
show how to construct amplitudes of the theory  by our approach.
Results in this section will be the same as those
 given in \cite{Feng:2009ei}. Here, the starting expression
 will be  obtained by the BCFW approach. Notice that the missing boundary
terms will be detected, although we do not pay attention to them.

\subsection{The six-point amplitude $A_6(1,2,3,4,5,6)$}

With only $\phi^4$ interaction, only amplitudes with even  number
of external particles can exist. The first  nontrivial
amplitude is  $A_6(1,2,3,4,5,6)$. Under the deformation
\bea
\la_1\to \la_1-z\la_2,~~~~\W\la_2\to \W\la_2+z\W\la_1,
\eea
there is only one pole $s_{561}$ detected and the corresponding
residue gives
\bea
A_0=-{1\over s_{561}},
\eea
which is our starting expression for the iterative construction.
Obviously, $A_0$ has the correct factorization limit for $s_{561}\to
0$.

The physical amplitude also contains poles $s_{123}$ and $s_{612}$,
for which $A_0$ cannot give the correct factorization limits. Under the
limit $s_{123}\to 0$, we have $\lim_{s_{123}\to 0} s_{123} A_6(1,2,3,4,5,6)=-1$,
but $\lim_{s_{123}\to 0} s_{123} A_0=0$, thus we need to add
${-1\over s_{123}}$ to $A_0$ to get the expression $\left({-1\over
s_{561}}+{-1\over s_{123}}\right)$ at the second step. Now it has
the correct factorization limits for poles $s_{561}$ and $s_{123}$, but
not for the pole $s_{612}$. Analogously, we add
a new term ${-1\over s_{612}}$ to get
\bea A_6(1,2,3,4,5,6)=-\Big({1\over s_{561}} +{1\over
s_{612}}+{1\over s_{123}}\Big).~~\label{phi4-6} \eea
Since all factorization limits of possible channels have been given correctly,
and no spurious pole appear (we will not emphasize the verification of the existence of spurious poles again if an expression does not contain any spurious pole), \eref{phi4-6} is the correct result. Although we did not try to find
the boundary term, the added terms in these steps give   the boundary
contribution  $\Big({-1\over s_{612}}+{-1\over s_{123}}\Big)$.
%

\subsection{The eight-point amplitude $A_8(1,2,3,4,5,6,7,8)$}

The second example is the eight-point amplitude
$A_8(1,2,3,4,5,6,7,8)$. Under the $\Spab{1|2}$-shift, the BCFW
approach gives
\bea
A_0=\Big[{1\over s_{781}}\Big({1\over s_{234}}+{1\over s_{345}}+{1\over s_{456}}\Big)
+{1\over s_{234}}\Big({1\over s_{567}}+{1\over s_{678}}\Big)\Big],
\eea
which gives correct factorization limits for poles $s_{781}$ and
$s_{234}$ detected by the deformation.

$A_0$ does not contain the pole $s_{123}$, which indicates
\bea
{\lim_{s_{123}\to 0}s_{123}A_0\over s_{123}}=0.
\eea
Hence when we consider the factorization limit for the pole $s_{123}$, a new term needs to be added
\bea  A_{s_{123}}=-{\lim_{s_{123}\to
0}A_4(1,2,3,-P_{123})A_6(P_{123},4,5,6,7,8)\over s_{123}}&=&{1\over
s_{123}}\Big({1\over s_{456}}+{1\over s_{567}}+{1\over
s_{678}}\Big). \eea
Thus we obtain $A_1=A_0+A_{s_{123}}$ at the second step, which gives
correct factorization limits for poles  $s_{123}$,
$s_{781}$ and $s_{234}$.

But $A_1$ does not contain  the pole $s_{812}$, similarly we need to add
\bea
A_{s_{812}}={1\over s_{812}}\Big({1\over s_{345}}+{1\over s_{456}}+{1\over s_{567}}\Big)
\eea
to get $A_2=A_1+A_{s_{812}}$, which has correct factorization
limits for  poles $s_{781}$, $s_{234}$, $s_{123}$ and $s_{812}$.

Next we consider the factorization limit for the pole $s_{678}\to 0$,
given by
\bea
A_{s_{678}}={1\over s_{678}}\Big({1\over s_{123}}+{1\over s_{234}}+{1\over s_{345}}\Big)~~\label{678}.
\eea
On the other hand, we have
\bea
\lim_{s_{678}\to 0}s_{678}A_2={1\over s_{123}}+{1\over s_{234}}~~\label{lim}.
\eea
Thus using the adjustment \eref{frame-A13}, we add the difference between \eref{678} and \eref{lim} to get
\bea
A_3=A_2+\Big(A_{s_{678}}-{\lim_{s_{678}\to 0}s_{678}A_2\over s_{678}}\Big)=A_2+{1\over s_{678}}{1\over s_{345}}.
\eea
It can be checked that $A_3$ provides correct factorization limits for
all possible channels, for instance,
\bea \lim_{s_{345}\to 0}s_{345}A_3={1\over s_{678}}+{1\over
s_{781}}+{1\over s_{812}}=- \lim_{s_{345}\to
0}A_L(3,4,5,-P_{345})A_R(P_{345},6,7,8,1,2). \eea
Therefore we have found the correct result
\bea
A_8(1,2,3,4,5,6,7,8)=A_3=\sum_{\sigma\in{Z_8}}\Big({1\over s_{\sigma(1)\sigma(2)\sigma(3)}s_{\sigma(6)\sigma(7)\sigma(8)}}
+{1\over 2s_{\sigma(1)\sigma(2)\sigma(3)}s_{\sigma(5)\sigma(6)\sigma(7)}}\Big),
\eea
where  the boundary term of $A_0$ is
\bea
B&=&A_{s_{123}}+A_{s_{812}}+{1\over s_{678}}{1\over s_{345}}\nn
&=&{1\over s_{123}}\Big({1\over s_{456}}+{1\over s_{567}}+{1\over s_{678}}\Big)+{1\over s_{812}}\Big({1\over s_{345}}+{1\over s_{456}}+{1\over s_{567}}\Big)+{1\over s_{678}}{1\over s_{345}}.
\eea
%

\subsection{The ten-point amplitude $A_{10}(1,2,3,4,5,6,7,8,9,10)$}

Now we consider the third example, the ten-point amplitude
$A_{10}(1,2,3,4,5,6,7,8,9,10)$. Using the $\Spab{1|2}$-shift, we
get the starting expression
\bea
A_0&=&{1\over s_{234}}\Big[{1\over s_{12345}}\Big({1\over s_{678}}+{1\over s_{789}}+{1\over s_{89(10)}}\Big)
+{1\over s_{56789}}\Big({1\over s_{567}}+{1\over s_{678}}+{1\over s_{789}}\Big)+{1\over s_{567}}\Big({1\over s_{89(10)}}+{1\over s_{9(10)1}}\Big)\Big]\nn
& &+{1\over s_{9(10)1}}\Big[{1\over s_{34567}}\Big({1\over s_{345}}+{1\over s_{456}}+{1\over s_{567}}\Big)
+{1\over s_{45678}}\Big({1\over s_{456}}+{1\over s_{567}}+{1\over s_{678}}\Big)+{1\over s_{678}}\Big({1\over s_{234}}+{1\over s_{345}}\Big)\Big]\nn
& &+{1\over s_{23456}}\Big({1\over s_{234}}+{1\over s_{345}}+{1\over s_{456}}\Big)\Big({1\over s_{789}}+{1\over s_{89(10)}}+{1\over s_{9(10)1}}\Big),
\eea
which has correct factorization limits for poles $s_{234}$,
$s_{9(10)1}$ and $s_{23456}$ detected by the deformation.

Since $A_0$ does not contain the pole $s_{123}$, we should add a term
to $A_0$ to provide the correct factorization limit. Similar
manipulations as previous lead to
$A_1=A_0+A_{s_{123}}$ where
\bea
A_{s_{123}}&=&{1\over s_{123}}\Big[{1\over s_{56789}}\Big({1\over s_{567}}+{1\over s_{678}}+{1\over s_{789}}\Big)
+{1\over s_{45678}}\Big({1\over s_{456}}+{1\over s_{567}}+{1\over s_{678}}\Big)+{1\over s_{12345}}\Big({1\over s_{678}}+{1\over s_{789}}+{1\over s_{89(10)}}\Big)\nn
& &+{1\over s_{456}}{1\over s_{789}}
+{1\over s_{456}}{1\over s_{89(10)}}+{1\over s_{567}}{1\over s_{89(10)}}\Big].
\eea
The new $A_1$ has correct factorization limits for  poles $s_{234}$,
$s_{9(10)1}$, $s_{23456}$ and $s_{123}$.

Now we move on to consider  the pole $s_{(10)12}$. After a little bit
computation, we get $A_2=A_1+A_{s_{(10)12}}$, where
\bea
A_{s_{(10)12}}&=&{1\over s_{(10)12}}\Big[{1\over s_{45678}}\Big({1\over s_{456}}+{1\over s_{567}}+{1\over s_{678}}\Big)
+{1\over s_{34567}}\Big({1\over s_{345}}+{1\over s_{456}}+{1\over s_{567}}\Big)+{1\over s_{56789}}\Big({1\over s_{567}}+{1\over s_{678}}+{1\over s_{789}}\Big)\nn
& &+{1\over s_{345}}{1\over s_{678}}+{1\over s_{345}}{1\over s_{789}}+{1\over s_{456}}{1\over s_{789}}\Big],
\eea
which provides correct factorization limits for poles $s_{234}$,
$s_{9(10)1}$, $s_{23456}$, $s_{123}$ and
$s_{(10)12}$.

Now we consider the pole $s_{345}$. The correct factorization limit is
\bea
A_{s_{345}}&=&{1\over s_{345}}\Big[{1\over s_{23456}}\Big({1\over s_{789}}+{1\over s_{89(10)}}+{1\over s_{9(10)1}}\Big)
+{1\over s_{12345}}\Big({1\over s_{678}}+{1\over s_{789}}+{1\over s_{89(10)}}\Big)\nn
& &+{1\over s_{34567}}\Big({1\over s_{89(10)}}+{1\over s_{9(10)1}}+{1\over s_{(10)12}}\Big)+{1\over s_{678}}{1\over s_{9(10)1}}
+{1\over s_{678}}{1\over s_{(10)12}}+{1\over s_{789}}{1\over s_{(10)12}}\Big],
\eea
while $A_2$ gives
\bea
\lim_{s_{345}\to 0}s_{345}A_2&=&{1\over s_{23456}}\Big({1\over s_{789}}+{1\over s_{89(10)}}+{1\over s_{9(10)1}}\Big)+{1\over s_{34567}}\Big({1\over s_{9(10)1}}+{1\over s_{(10)12}}\Big)\nn
& &+{1\over s_{678}}{1\over s_{9(10)1}}
+{1\over s_{678}}{1\over s_{(10)12}}+{1\over s_{789}}{1\over s_{(10)12}}.
\eea
Adding the difference, we can construct
\bea
A_3&=&A_2+\Big(A_{s_{345}}-{\lim_{s_{345}\to 0}s_{345}A_2\over s_{345}}\Big)\nn
&=&A_2+{1\over s_{345}}\Big[{1\over s_{12345}}\Big({1\over s_{678}}+{1\over s_{789}}+{1\over s_{89(10)}}\Big)+{1\over s_{89(10)}}{1\over s_{34567}}\Big].
\eea
Then $A_3$ provides correct factorization limits for poles $s_{234}$,
$s_{9(10)1}$, $s_{23456}$, $s_{123}$, $s_{(10)12}$ and $s_{345}$.

Finally  we consider the pole $s_{89(10)}$, whose correct factorization
limit is
\bea
A_{s_{89(10)}}&=&{1\over s_{89(10)}}\Big[{1\over
s_{23456}}\Big({1\over s_{234}}+{1\over s_{345}}+{1\over
s_{456}}\Big) +{1\over s_{12345}}\Big({1\over s_{123}}+{1\over
s_{234}}+{1\over s_{345}}\Big)\nn & &+{1\over s_{34567}}\Big({1\over
s_{345}}+{1\over s_{456}}+{1\over s_{567}}\Big)+{1\over
s_{123}}{1\over s_{456}} +{1\over s_{123}}{1\over s_{567}}+{1\over
s_{234}}{1\over s_{567}}\Big],\eea
while $A_3$ gives
\bea   \lim_{s_{89(10)}\to 0}s_{89(10)}A_3&=&{1\over
s_{23456}}\Big({1\over s_{234}}+{1\over s_{345}}+{1\over
s_{456}}\Big) +{1\over s_{12345}}\Big({1\over s_{123}}+{1\over
s_{234}}+{1\over s_{345}}\Big)\nn & &+{1\over s_{34567}}{1\over
s_{345}}+{1\over s_{123}}{1\over s_{456}} +{1\over s_{123}}{1\over
s_{567}}+{1\over s_{234}}{1\over s_{567}}, \eea
thus we can construct
\bea
A_4&=&A_3+\Big(A_{s_{89(10)}}-{\lim_{s_{89(10)}\to 0}s_{89(10)}A_3\over s_{89(10)}}\Big)\nn
&=&A_3+{1\over s_{89(10)}}\Big[{1\over s_{34567}}\Big({1\over s_{456}}+{1\over s_{567}}\Big)\Big].
\eea
One can verify that $A_4$ gives correct factorization limits for all
channels. Hence, we have found the final result
\bea
A_4&=&\sum_{\sigma\in Z_{10}}\Big({1\over s_{\sigma(1)\sigma(2)\sigma(3)}s_{\sigma(1)\sigma(2)\sigma(3)\sigma(4)\sigma(5)}s_{\sigma(8)\sigma(9)\sigma(10)}}+{1\over s_{\sigma(1)\sigma(2)\sigma(3)}s_{\sigma(1)\sigma(2)\sigma(3)\sigma(4)\sigma(5)}s_{\sigma(7)\sigma(8)\sigma(9)}}\nn
& &+{1\over s_{\sigma(1)\sigma(2)\sigma(3)}s_{\sigma(10)\sigma(1)\sigma(2)\sigma(3)\sigma(4)}s_{\sigma(7)\sigma(8)\sigma(9)}}
+{1\over 2s_{\sigma(1)\sigma(2)\sigma(3)}s_{\sigma(10)\sigma(1)\sigma(2)\sigma(3)\sigma(4)}s_{\sigma(6)\sigma(7)\sigma(8)}}\nn
& &+{1\over 2s_{\sigma(1)\sigma(2)\sigma(3)}s_{\sigma(1)\sigma(2)\sigma(3)\sigma(4)\sigma(5)}s_{\sigma(6)\sigma(7)\sigma(8)}}
+{1\over 2s_{\sigma(1)\sigma(2)\sigma(3)}s_{\sigma(9)\sigma(10)\sigma(1)\sigma(2)\sigma(3)}s_{\sigma(6)\sigma(7)\sigma(8)}}\nn
& &+{1\over s_{\sigma(1)\sigma(2)\sigma(3)}s_{\sigma(4)\sigma(5)\sigma(6)}s_{\sigma(7)\sigma(8)\sigma(9)}}\Big).
\eea
As a byproduct, the boundary term of $A_0$ is
\bea
B=A_{s_{123}}+A_{s_{(10)12}}+{1\over s_{345}}{1\over s_{12345}}\Big({1\over s_{678}}+{1\over s_{789}}+{1\over s_{89(10)}}\Big)
+{1\over s_{89(10)}}{1\over s_{34567}}\Big({1\over s_{345}}+{1\over s_{456}}+{1\over s_{567}}\Big).
\eea
%

\section{Example 2: Pure gauge theory \label{4}}

Now we move on to color ordered amplitudes of gluons. The
lowest-point amplitudes are three-point MHV and anti-MHV amplitudes,
which are given as
\bea
A_3(1^-,2^-,3^+)={\Spaa{1|2}^4\over\Spaa{1|2}\Spaa{2|3}\Spaa{3|1}},~~~~A_3(1^+,2^+,3^-)={\Spbb{1|2}^4\over\Spbb{1|2}\Spbb{2|3}\Spbb{3|1}},
\eea
where the coupling constant has been neglected. As well known,
these amplitudes will vanish when $z\to \infty$ under correct
deformations, therefore they can be computed by the BCFW
approach \cite{ArkaniHamed:2008yf}. We will use our approach to reproduce
them. Results in this section can also be found
in \cite{Feng:2011np}. In this section, the calculation will start by
computing the factorization limit for one channel.

\subsection{The MHV amplitude $A_n(1^+...i^-...j^-...n^+)$}

The first case is the $n$-point MHV amplitude, given by the well
known formula
\bea
A_n(1^+\cdots i^-\cdots j^-\cdots n^+)={\Spaa{i|j}^4\over \Spaa{1|2}\Spaa{2|3}\cdots\Spaa{n-1|n}\Spaa{n|1}}.~~\label{MHV}
\eea
It is sufficient to consider $A_n(1^-\cdots
i^-\cdots n^+)$ since the  general formula can be
transformed into this choice by cyclic permutation. We assume \eref{MHV} is
valid for $m$-point MHV amplitudes with $m<n$, then consider
factorization limits of the $n$-point MHV amplitude. First, let us
consider the limit $s_{12}\to 0$. There are two types of solutions
\bea &
&I_1:~~\la_2=\alpha\la_1,~~P_{12}=\la_1(\W\la_1+\alpha\W\la_2),\nn &
&I_2:~~\W\la_2=\beta\W\la_1,~~P_{12}=(\la_1+\beta\la_2)\W\la_1. \eea
Solution $I_2$ contributes nothing to the factorization limit,
since no matter which helicity is assigned for the internal propagator, one
of the sub-amplitudes $A_L$ and $A_R$ vanishes, thus only solution
$I_1$ is considered.  Then$^{[4]}$ \footnotetext[4]{ For the complex
momentum $-P$, one can choose corresponding spinors as
$\la_{-P}=\la_P$ and $\W\la_{-P}=-\W\la_P$. In this choice,
$\W\la_{-P}$ can be replaced by $\W\la_P$ if it appears for even times.}
\bea
\lim_{\Spaa{1|2}\to 0}A_3(1^-,2^+,-P_{12}^+)A_{n-1}(P_{12}^-,3^+,\cdots i^-\cdots n^+)
&=&{\Spbb{2|-P_{12}}^3\over\Spbb{-P_{12}|1}\Spbb{1|2}}{\Spaa{i|P_{12}}^4\over\Spaa{P|3}\Spaa{3|4}\cdots\Spaa{n|P_{12}}}\nn
&=&{\Spbb{2|1}^3\over\alpha\Spbb{2|1}\Spbb{1|2}}{\Spaa{i|1}^4\over\Spaa{1|3}\Spaa{3|4}\cdots\Spaa{n|1}}\nn
&=&{\Spbb{1|2}\Spaa{1|i}^4\over\Spaa{2|3}\Spaa{3|4}\cdots\Spaa{n|1}},
\eea
where we have used $\a\Spaa{1|3}=\Spaa{2|3}$. From
this we can get the starting expression
\bea
A_{\Spaa{1|2}}=-{\lim_{\Spaa{1|2}\to 0}A_LA_R\over s_{12}}={\Spaa{1|i}^4\over\Spaa{1|2}\Spaa{2|3}\Spaa{3|4}\cdots\Spaa{n|1}}.
\eea
Although for this special case it is already the correct result,
logically, we still need to check whether it has correct
factorization limits for other channels. For instance, let us
consider the limit $s_{(j-1)j}\to 0$ where both $(j-1)$ and $j$ have
positive helicity. The non-vanishing sub-amplitude corresponds to the
solution $\la_j=\alpha\la_{j-1},~P_{(j-1)j}=\la_{j-1}(\W\la_{j-1}+\alpha\W\la_j)$. Then we have
\bea
& &\lim_{\Spaa{j-1|j}\to 0}A_3((j-1)^+,j^+,-P_{(j-1)j}^-)A_{n-1}(P_{(j-1)j}^+,(j+1)^+,\cdots1^-\cdots i^-\cdots(j-2)^+)\nn
&=&{\Spbb{j-1|j}^3\over\Spbb{j|-P_{(j-1)j}}\Spbb{-P_{(j-1)j}|j-1}}
{\Spaa{1|i}^4\over\Spaa{P_{(j-1)j}|j+1}\Spaa{j+1|j+2}\cdots\Spaa{j-2|P_{(j-1)j}}}\nn
&=&
{\Spbb{j-1|j}\Spaa{1|i}^4\over\alpha\Spaa{j-1|j+1}\Spaa{j+1|j+2}\cdots\Spaa{j-2|j-1}}\nn
&=&{\Spbb{j-1|j}\Spaa{1|i}^4\over\Spaa{j|j+1}\Spaa{j+1|j+2}\cdots\Spaa{j-2|j-1}}\nn
&=&\lim_{\Spaa{j-1|j}\to 0}s_{(j-1)j}A_{\Spaa{1|2}},
\eea
therefore $A_{\Spaa{1|2}}$ provides the correct factorization limit for $s_{(j-1)j}\to 0$. It
can also be checked that $A_{\Spaa{1|2}}$ provides correct factorization limits for other channels. Therefore we can conclude that $A_{\Spaa{1|2}}$ is the amplitude $A_n(1^-\cdots i^-\cdots n^+)$
that has all correct factorization limits.

\subsection{The six-point amplitude $A_6(1^-,2^-,3^-,4^+,5^+,6^+)$}

Now we turn to the six-point NMHV amplitude
$A_6(1^-,2^-,3^-,4^+,5^+,6^+)$.  First let's consider the limit
$s_{12}\to 0$. The solution for the non-vanishing sub-amplitude is
\bea
\W\la_2=\alpha\W\la_1,~~P_{12}=(\la_1+\alpha\la_2)\W\la_1,~~~~\label{solu}
\eea
We use an auxiliary spinor $\eta$ to express the un-determined
parameter $\alpha$  as $\alpha={\Spbb{\eta|2}\over\Spbb{\eta|1}}$.
Then
\bea
A_{\Spbb{1|2}}&=&-{\lim_{\Spbb{1|2}\to 0}A_3(1^-,2^-,-P_{12}^+)A_5(P_{12}^-,3^-,4^+,5^+,6^+)\over s_{12}}\nn
&=&{1\over\Spbb{1|2}\Spbb{\eta|1}\Spbb{2|\eta}}{\Spab{3|1+2|\eta}^3\over \Spaa{3|4}\Spaa{4|5}\Spaa{5|6}\Spab{6|1+2|\eta}}.
\eea
Notice that the spinor $\eta$ can be chosen arbitrarily. It is exactly
the ambiguity we have emphasized in section 2. Different
choices of $\eta$ gives the same result only under the limit
$\Spbb{1|2}\to 0$. Also since there are three pairs of $\eta$ (we
count one in the numerator and one in the denominator as a pair), each pair
can be chosen independently. For the current example, we choose three pairs
of $\eta$ to be the same. In other words, we have chosen a type of
representative expressions in the category of the limit $\Spbb{1|2}\to
0$.

To fix $\eta$, we can try to choose one value so that $A_{\Spbb{1|2}}$ has correct factorization limits
for other channels, thus we pick a pole contained in
$A_{\Spbb{12}}$. Let us consider the limit $s_{45}\to 0$, the
solution corresponding to the non-vanishing sub-amplitudes is
\bea
\la_5=\alpha\la_4,~~P_{45}=\la_4(\W\la_4+\alpha\W\la_5),~~\alpha={\Spaa{\zeta|4}\over\Spaa{\zeta|5}},
\eea
then we have
\bea
A_{\Spaa{4|5}}&=&-{\lim_{\Spaa{4|5}\to 0}A_3(4^+,5^+,-P_{45}^-)A_5(P_{45}^+,6^+,1^-,2^-,3^-)\over s_{45}}\nn
&=&{1\over\Spaa{4|5}\Spaa{\zeta|4}\Spaa{5|\zeta}}{\Spab{\zeta|4+5|6}^3\over\Spbb{6|1}\Spbb{1|2}\Spbb{2|3}\Spab{\zeta|4+5|3}}.
\eea
Again, the form of $A_{\Spaa{4|5}}$ provides a type of representative
expressions in the category of the limit $\Spaa{4|5}\to 0$. Now we ask if there is a choice of $\eta$ and $\zeta$, such
that above two representative expressions are the same under
corresponding limits, \textit{i.e.},
$\left(s_{12}s_{45}A_{\Spbb{1|2}}\right)_{\Spbb{1|2}\to
0}=\left(s_{12}s_{45}A_{\Spaa{4|5}}\right)_{\Spaa{4|5}\to 0}$. This
is a strong constraint, since it means that two correct factorization limits
are given either by $A_{\Spbb{1|2}}$ or by $A_{\Spaa{4|5}}$. In general, it cannot be achieved, but for this
case, we fortunately manage to obtain the choice
$\eta=\W\la_6,~~\zeta=\la_3$. To check it, for the left hand side,
we have
\bea
\lim_{\Spbb{1|2}\to 0}A_3(1^-,2^-,-P_{12}^+)A_5(P_{12}^-,3^-,4^+,5^+,6^+)&=&{\Spaa{1|2}\Spab{3|1+2|6}^3\over\Spaa{3|4}\Spaa{4|5}\Spaa{5|6}\Spbb{6|1}\Spbb{2|6}\Spab{6|1+2|6}}\nn
&=&{\Spaa{1|2}\Spab{3|4+5|6}^3\over\Spaa{3|4}\Spaa{4|5}\Spbb{6|1}\Spab{5|6|2}(s_{16}+s_{26})}\nn
&=&{\Spaa{1|2}\Spab{3|4+5|6}^3\over\Spaa{3|4}\Spaa{4|5}\Spbb{1|6}\Spab{5|3+4|2}s_{126}},
\eea
where $\Spbb{1|2}=0$ is used in the last step, thus
$s_{16}+s_{26}=s_{12}+s_{16}+s_{26}=s_{126}$ and
$\Spab{5|3+4|2}=-\Spab{5|6|2}$. Similarly, for the right hand side
we have
\bea
\lim_{\Spaa{4|5}\to 0}A_3(4^+,5^+,-P_{45}^-)A_5(P_{45}^+,6^+,1^-,2^-,3^-)&=&{\Spbb{4|5}\Spab{3|4+5|6}^3\over\Spaa{3|4}\Spbb{1|2}\Spbb{1|6}\Spab{5|3+4|2}s_{126}},
\eea
where we have used $\Spab{3|4+5|3}=s_{126}$ and
$\Spaa{53}\Spbb{2|3}=-\Spab{5|3+4|2}$ under the limit $\Spaa{4|5}\to
0$.

Now a nice starting expression appears
\bea
A_{\Spbb{1|2}\Spaa{4|5}}=A_{\Spbb{1|2}}=A_{\Spaa{4|5}}=-{\Spab{3|4+5|6}^3\over\Spaa{3|4}\Spaa{4|5}\Spbb{1|2}\Spbb{6|1}\Spab{5|3+4|2}s_{126}},
\eea
To continue, we consider other poles. One nice choice is a pole
$s$ such that $\lim_{s\to 0}s A_{\Spbb{1|2}\Spaa{4|5}}=0$. There are
two two-particle channels $s_{23}\to 0$ and $s_{56}\to 0$ satisfying this requirement.
Proceeding as the case $\Spbb{1|2}\to 0$ and $\Spaa{4|5}\to 0$, we get
\bea
A_{\Spbb{2|3}\Spaa{5|6}}=-{\Spab{1|2+3|4}^3\over\Spaa{6|1}\Spaa{5|6}\Spbb{2|3}
\Spbb{3|4}\Spab{5|3+4|2}s_{234}},
\eea
which has correct factorization limits for $s_{23}\to 0$ and
$s_{56}\to 0$. Since $A_{\Spbb{1|2}\Spaa{4|5}}$ and
$A_{\Spbb{2|3}\Spaa{5|6}}$ do not share any physical pole, we should
sum them to get
\bea
A_{\Spbb{1|2}\Spbb{2|3}\Spaa{4|5}\Spaa{5|6}}&=&A_{\Spbb{1|2}\Spaa{4|5}}+A_{\Spbb{2|3}\Spaa{5|6}}\nn
&=&-{1\over\Spab{5|3+4|2}}
\Big({\Spab{1|2+3|4}^3\over\Spaa{6|1}\Spaa{5|6}\Spbb{2|3}\Spbb{3|4}s_{234}}
+{\Spab{3|4+5|6}^3\over\Spaa{3|4}\Spaa{4|5}\Spbb{1|2}\Spbb{6|1}s_{126}}\Big),~~~~\label{glu-A61}
\eea
which gives the correct  factorization limits for $s_{12}\to 0$,
$s_{23}\to 0$, $s_{45}\to 0$ and $s_{56}\to 0$. One can verify that
$A_{\Spbb{1|2}\Spbb{2|3}\Spaa{4|5}\Spaa{5|6}}$ also gives
correct factorization limits for remaining channels$^{[5]}$\footnotetext[5]{There is a
technical issue regarding the limit such as $s_{234}\to 0$. For
this case, the spinor $\la_{P_{234}}$ can be expressed, for example, via
$\Spaa{a|P_{234}}= {\Spab{a|P_{234}|b}\over \Spbb{P_{234}|b}}$ (see in Appendix B). }.

One can observe the factor $\Spab{5|3+4|2}$ in the denominator of $A_{\Spbb{1|2}\Spbb{2|3}\Spaa{4|5}\Spaa{5|6}}$. However,
it is not a pole since
\bea
\lim_{\Spab{5|3+4|2}\to 0}\Spab{5|3+4|2}A_{\Spbb{1|2}\Spbb{2|3}\Spaa{4|5}\Spaa{5|6}}=0.~~~~\label{veri}
\eea
To verify this, notice that $\Spab{5|3+4|2}\to 0$ implies $\ket{5}\propto |3+4\bket{2}$. Then the momentum conservation condition
becomes
\bea
\ket{1}\bbra{1}+\ket{2}\bbra{2}+\ket{3}\bbra{3}+\ket{4}\bbra{4}+c|3+4\bket{2}\bbra{5}+\ket{6}\bbra{6}=0.~~~~\label{conserv}
\eea
The coefficient $c$ can be fixed by contract \eref{conserv} with two spinors. For example, contracting \eref{conserv} with
$\bket{1}$ and $\bra{6}$, we get $c=-{\Spab{1|2+3+4|6}\over \Spab{1|3+4|2}\Spbb{5|6}}$. Contracting with different spinors gives different expressions of $c$ but they are equivalent under the limit $\Spab{5|3+4|2}\to 0$.$^{[6]}$\footnotetext[6]{If one use this way to fix the coefficient $\alpha$ in \eref{solu},
contracting the momentum conservation equation with different spinors indeed provides different choices of the reference spinor $\eta$.
There is no guidance to show which choice is more proper for latter calculation.} Substituting these into \eref{glu-A61},
one can get the result \eref{veri}. Consequently, the expression \eref{glu-A61} contains only physical poles. Thus, we have found
\bea
A_6(1^-,2^-,3^-,4^+,5^+,6^+)=A_{\Spbb{1|2}\Spbb{2|3}\Spaa{4|5}\Spaa{5|6}}.
\eea

Although in this subsection, we start with the factorization of
a two-particle channel, one can start with the factorization of
a three-particle channel and proceed similarly to get the correct result.
The calculation is shown in Appendix B.

\subsection{The six-point amplitude $A_6(1^+,2^-,3^+,4^-,5^+,6^-)$}

Let us start with the factorization limit for $s_{12}\to 0$. There are two
types of solutions for non-vanishing sub-amplitudes:
\bea
& &I_1:~~\la_2=\alpha_1\la_1,~~P_{12}=\la_1(\W\la_1+\alpha_1\W\la_2),~~\alpha_1={\Spaa{\eta_1| 2}\over\Spaa{\eta_1| 1}},\nn
& &I_2:~~\W\la_2=\beta_1\W\la_1,~~P_{12}=(\la_1+\beta_1\la_2)\W\la_1,~~\beta_1={\Spbb{\zeta_1| 2}\over\Spbb{\zeta_1| 1}}.
\eea
For solution $I_1$, we have
\bea
A_{\Spaa{1|2}}={\Spaa{2|\eta_1}^3\over\Spaa{\eta_1|
1}\Spaa{1|2}}{\Spbb{3|5}^4\over\Spbb{3|4}\Spbb{4|5}\Spbb{5|6}\Spab{\eta_1|1+2|6}
\Spab{\eta_1|1+2|3}}, \eea
and for solution $I_2$,
\bea
A_{\Spbb{1|2}}={\Spbb{1|\zeta_1}^3\Spaa{4|6}^4\over\Spbb{2|\zeta_1}\Spbb{2|1}\Spaa{3|4}\Spaa{4|5}\Spaa{5|6}\Spab{6|1+2|\zeta_1}
\Spab{3|1+2|\zeta_1}}. \eea
$A_{\Spaa{1|2}}$ does not contain the pole  $\Spbb{1|2}$ and
$A_{\Spbb{1|2}}$ does not contain the pole $\Spaa{1|2}$. Thus we
sum these two parts to obtain the starting expression
$A_{\Spaa{1|2}\Spbb{1|2}}=A_{\Spaa{1|2}}+A_{\Spbb{1|2}}$, which
satisfies the factorization limit for $s_{12}\to 0$.

In the expression of $A_{\Spaa{1|2}\Spbb{1|2}}$, there are unfixed
variables $\zeta_1$ and $\eta_1$, which reflects the freedom of
representative elements in the category as discussed in section 2.
Now we try to fix these parameters as previous. To
do so, consider the limit $s_{23}\to 0$,  similar calculation
leads to
\bea
s_{23}\to 0:& &A_{\Spaa{2|3}}={\Spaa{2|\eta_2}^3\Spbb{5|1}^4\over\Spaa{3|\eta_2}\Spaa{3|2}\Spbb{4|5}\Spbb{5|6}\Spbb{6|1}\Spab{\eta_2|2+3|1}
\Spab{\eta_2|2+3|4}},\nn
& &A_{\Spbb{2|3}}={\Spbb{3|\zeta_2}^3\Spaa{4|6}^4\over\Spbb{\zeta_2|2}\Spbb{3|2}\Spaa{4|5}\Spaa{5|6}\Spaa{6|1}\Spab{1|2+3|\zeta_2}
\Spab{4|2+3|\zeta_2}}.
\eea
Matching $A_{\Spbb{2|3}}$ with $A_{\Spbb{1|2}}$, yields
\bea
\zeta_1=\W\la_3,~~\zeta_2=\W\la_1,
\eea
and
\bea
A_{\Spbb{1|2}\Spbb{2|3}}=A_{\Spbb{1|2}}=A_{\Spbb{2|3}}={\Spbb{1|3}^4\Spaa{4|6}^4\over\Spaa{4|5}\Spaa{5|6}\Spbb{1|2}\Spbb{2|3}
\Spab{6|1+2|3}\Spab{4|2+3|1}s_{123}}.~~\label{part 1}
\eea
Plugging this back, the staring expression becomes
$A_{\Spaa{1|2}\Spbb{1|2}\Spbb{2|3}}=A_{\Spaa{1|2}}+A_{\Spbb{1|2}\Spbb{2|3}}$ which
has correct factorization limits for $s_{12}\to 0$ plus
$\Spbb{2|3}\to 0$. When we do the iterative step to include the new
factorization limit $\Spaa{2|3}\to 0$, we need to add $A_{\Spaa{2|3}}$
to reach  $A_{\Spaa{1|2}\Spaa{2|3}\Spbb{1|2}\Spbb{2|3}}=
A_{\Spaa{1|2}}+A_{\Spaa{2|3}}+A_{\Spbb{1|2}\Spbb{2|3}}$, which has correct
factorization limits for $s_{12}\to 0$ and $s_{23}\to 0$.

Now we continue to include new factorization limits for $s_{34}\to 0$.
Using
\bea
s_{34}\to 0:& &A_{\Spaa{3|4}}={\Spaa{4|\eta_3}^3\over\Spaa{\eta_3|3}\Spaa{3|4}}{\Spbb{5|1}^4\over\Spbb{5|6}\Spbb{6|1}\Spbb{1|2}\Spab{\eta_3|3+4|2}
\Spab{\eta_3|3+4|5}},\nn
& &A_{\Spbb{3|4}}={\Spbb{3|\zeta_3}^3\Spaa{6|2}^4\over\Spbb{4|\zeta_3}\Spbb{4|3}\Spaa{5|6}\Spaa{6|1}\Spaa{1|2}\Spab{2|3+4|\zeta_3}
\Spab{5|3+4|\zeta_3}},
\eea
we find that the freedom of $A_{\Spaa{2|3}}$ can be fixed by
$\eta_2=\la_4$ when we match it with $A_{\Spaa{3|4}}$ with
$\eta_3=\la_2$, and this single expression is
\bea
A_{\Spaa{2|3}\Spaa{3|4}}=A_{\Spaa{2|3}}=A_{\Spaa{3|4}}={\Spaa{2|4}^4\Spbb{5|1}^4
\over\Spaa{2|3}\Spaa{3|4}\Spbb{5|6}\Spbb{6|1}\Spab{4|2+3|1}\Spab{2|3+4|5}s_{234}}.~~\label{part 2}
\eea
Similarly, matching $A_{\Spbb{3|4}}$ and $A_{\Spaa{1|2}}$ we find
$\eta_1=\la_6,~~\zeta_3=\W\la_5$ and
\bea
A_{\Spaa{1|2}\Spbb{3|4}}=A_{\Spaa{1|2}}=A_{\Spbb{3|4}}={\Spaa{2|6}^4\Spbb{3|5}^4\over\Spaa{6|1}
\Spaa{1|2}\Spbb{3|4}\Spbb{4|5}\Spab{6|4+5|3}\Spab{2|3+4|5}s_{345}}.~~\label{part 3}
\eea
Summing \eref{part 1}, \eref{part 2} and \eref{part 3} we have
\bea
A_{\Spaa{1|2}\Spaa{2|3}\Spaa{3|4}\Spbb{1|2}\Spbb{2|3}\Spbb{3|4}}&=&A_{\Spbb{1|2}\Spbb{2|3}}+A_{\Spaa{2|3}\Spaa{3|4}}+A_{\Spaa{1|2}\Spbb{3|4}}\nn
&=&{\Spbb{1|3}^4\Spaa{4|6}^4\over\Spbb{1|2}\Spbb{2|3}\Spaa{4|5}\Spaa{5|6}\Spab{6|1+2|3}\Spab{4|2+3|1}s_{123}}\nn
& &+{\Spaa{2|4}^4\Spbb{5|1}^4
\over\Spaa{2|3}\Spaa{3|4}\Spbb{5|6}\Spbb{6|1}\Spab{4|2+3|1}\Spab{2|3+4|5}s_{234}}\nn
& &+{\Spaa{2|6}^4\Spbb{3|5}^4\over\Spaa{6|1}
\Spaa{1|2}\Spbb{3|4}\Spbb{4|5}\Spab{6|4+5|3}\Spab{2|3+4|5}s_{345}}.
\eea
One can check that $A_{\Spaa{1|2}\Spaa{2|3}\Spaa{3|4}\Spbb{1|2}\Spbb{2|3}\Spbb{3|4}}$ has correct factorization limits for all channels.
Again, it can be verified that there is no spurious pole. Thus we find
\bea
A_6(1^+,2^-,3^+,4^-,5^+,6^-)=A_{\Spaa{1|2}\Spaa{2|3}\Spaa{3|4}\Spbb{1|2}\Spbb{2|3}\Spbb{3|4}}.
\eea
%

\subsection{The six-point amplitude $A_6(1^+,2^+,3^-,4^+,5^-,6^-)$}

Let us start with factorization limit of  $s_{12}\to 0$. There is
only one type of solution corresponding to the non-vanishing
sub-amplitude, namely $ \la_2=\alpha\la_1$, which leads to
\bea
A_{\Spaa{1|2}}={\Spab{\eta_1|1+2|4}^4\over\Spaa{2|1}\Spaa{2|\eta_1}\Spaa{1|\eta_1}
\Spbb{3|4}\Spbb{4|5}\Spbb{5|6}\Spab{\eta_1|1+2|3}\Spab{\eta_1|1+2|6}}.
\eea
To fix the choice of $\eta_1$, noticing that the pole $\Spbb{5|6}$ in
$A_{\Spaa{1|2}}$, we then match it with the factorization limit
\bea
A_{\Spbb{5|6}}={\Spab{3|5+6|\zeta_1}^4\over\Spbb{6|5}\Spbb{\zeta_1|5}\Spbb{\zeta_1|6}\Spaa{1|2}\Spaa{2|3}\Spaa{3|4}\Spab{1|5+6|\zeta_1}
\Spab{4|5+6|\zeta_1}},
\eea
thus  we find $\eta_1=\la_3,~~\zeta_1=\W\la_4$ and the starting
expression given by
\bea &
&A_{\Spaa{1|2}\Spbb{5|6}}=A_{\Spaa{1|2}}=A_{\Spbb{5|6}}=-{\Spab{3|1+2|4}^4\over\Spaa{1|2}\Spaa{2|3}
\Spbb{4|5}\Spbb{5|6}\Spab{1|2+3|4}\Spab{3|1+2|6}s_{123}}.
\eea
Since $A_{\Spaa{1|2}\Spbb{5|6}}$ does not contain poles  $s_{34}$ and
$s_{61}$, we will try to include their factorization limits. The
limit  $s_{34}\to 0$ gives
\bea
& &A_{\Spaa{3|4}}={\Spaa{3|\eta_2}^3\Spbb{1|2}^3\over\Spaa{4|3}\Spaa{4|\eta_2}\Spbb{5|6}\Spbb{6|1}\Spab{\eta_2|3+4|2}
\Spab{\eta_2|3+4|5}},\nn
& &A_{\Spbb{3|4}}={\Spaa{5|6}^3\Spbb{\zeta_2| 4}^3\over \Spbb{4|3}\Spbb{\zeta_2 |3}\Spaa{6|1}\Spaa{1|2}
\Spab{2|3+4|\zeta_2}\Spab{5|3+4|\zeta_2}},
\eea
while the limit $s_{61}\to 0$ gives
\bea
& &A_{\Spaa{6|1}}={\Spaa{\eta_3 |6}^3\Spbb{2|4}^4\over\Spaa{1|6}\Spaa{1|\eta_3}\Spbb{2|3}\Spbb{3|4}\Spbb{4|5}\Spab{\eta_3|1+6|5}\Spab{\eta_3|1+6|2}},\nn
& &A_{\Spbb{6|1}}={\Spaa{3|5}^4\Spbb{1|\zeta_3}^4\over\Spbb{1|6}\Spbb{6|\zeta_3}\Spaa{2|3}\Spaa{3|4}\Spaa{4|5}
\Spab{5|6+1|\zeta_3}\Spab{2|6+1|\zeta_3}}.
\eea
Matching $A_{\Spbb{6|1}}$ with $A_{\Spaa{3|4}}$ fixes
$\eta_2=\la_5,~~\zeta_3=\W\la_2$, and we get
\bea
A_{\Spaa{3|4}\Spbb{6|1}}=A_{\Spaa{3|4}}=A_{\Spbb{6|1}}=-{\Spaa{3|5}^4\Spbb{1|2}^3\over\Spaa{3|4}\Spaa{4|5}\Spbb{6|1}
\Spab{5|3+4|2}\Spab{3|4+5|6}s_{345}}. \eea
Then matching $A_{\Spaa{6|1}}$ with $A_{\Spbb{3|4}}$ fixes
$\eta_3=\la_5,~~\zeta_2=\W\la_2$ and
\bea
A_{\Spaa{6|1}\Spbb{3|4}}=A_{\Spaa{6|1}}=A_{\Spbb{3|4}}=-{\Spaa{5|6}^3\Spbb{2|4}^4\over
\Spaa{6|1}\Spbb{2|3}\Spbb{3|4}\Spab{1|2+3|4} \Spab{5|3+4|2}s_{234}}.
\eea
Summing these, we have
\bea
A_{\Spaa{1|2}\Spaa{3|4}\Spaa{6|1}\Spbb{3|4}\Spbb{5|6}\Spbb{6|1}}&=&A_{\Spaa{1|2}\Spbb{5|6}}+A_{\Spaa{3|4}\Spbb{6|1}}+A_{\Spaa{6|1}\Spbb{3|4}}\nn
&=&-{\Spab{3|1+2|4}^4\over\Spaa{1|2}\Spaa{2|3}\Spbb{4|5}\Spbb{5|6}\Spab{1|2+3|4}\Spab{3|1+2|6}s_{123}}\nn
& &-{\Spaa{3|5}^4\Spbb{1|2}^3\over\Spbb{6|1}\Spaa{3|4}\Spaa{4|5}
\Spab{5|3+4|2}\Spab{3|4+5|6}s_{345}}\nn
& &-{\Spaa{5|6}^3\Spbb{2|4}^4\over \Spbb{2|3}\Spbb{3|4}\Spaa{6|1}\Spab{1|2+3|4}
\Spab{5|3+4|2}s_{234}}.
\eea
It can be verified that $A_{\Spaa{1|2}\Spaa{3|4}\Spaa{6|1}\Spbb{3|4}\Spbb{5|6}\Spbb{6|1}}$ has correct factorization limits for all channels,
and all spurious poles are canceled, therefore
\bea
A_6(1^+,2^+,3^-,4^+,5^-,6^-)=A_{\Spaa{1|2}\Spaa{3|4}\Spaa{6|1}\Spbb{3|4}\Spbb{5|6}\Spbb{6|1}}.
\eea
%

\section{Example 3: Einstein-Maxwell theory \label{5}}

In this section we consider amplitudes of photons coupling to
gravitons. In such a theory the lowest-point amplitudes are of
two types: photon-photon-graviton and three-graviton
self interaction
\bea
& &A_3(a_\gamma^{-1},b_\gamma^{+1},c_g^{-2})=\kappa{\Spaa{c|a}^4\over\Spaa{a|b}^2},~~~~~~~~
A_3(a_\gamma^{-1},b_\gamma^{+1},c_g^{+2})=\kappa{\Spbb{b|c}^4\over\Spbb{a|b}^2},\nn
& &A_3(a_g^{-2},b_g^{-2},c_g^{+2})=\kappa {\Spaa{a|b}^8\over\Spaa{a|b}^2\Spaa{b|c}^2\Spaa{c|a}^2},~~~~~~~~
A_3(a_g^{+2},b_g^{+2},c_g^{-2})=\kappa {\Spbb{a|b}^8\over\Spbb{a|b}^2\Spbb{b|c}^2\Spbb{c|a}^2}.
\eea
Since two types of three-point amplitudes have the same coupling
constant  $\kappa$, we will neglect $\kappa$ from now on.

Before starting the calculation, let us give a brief review on
some general  properties of amplitudes in this  theory. First, let us
review the validity of the BCFW approach. As proved by Cheung, for amplitudes contain at least one graviton, the
boundary term is zero under some proper deformations \cite{Cheung:2008dn}. Hence, the BCFW approach is available for
amplitudes containing gravitons. On the other hand, Arkani-Hamed and Kaplan have
shown that the boundary term will not vanish when deforming two
photons \cite{ArkaniHamed:2008yf}. However, their conclusion cannot
be applied to the circumstance that two deformed photons have the same
helicity. The reason is, their approach relies on the picture that a
hard photon moves in a soft background, which means two deformed
photons should be connected by photon propagators or they are attached  to
the same vertex. In Einstein-Maxwell theory, two photons
with the same helicity could not satisfy such a condition because of
the helicity structure of three-point amplitudes. In some cases, the
naive power counting of individual Feynman diagrams shows $A(z\to
\infty)=0$ when deforming two photons with the same helicity, for
example the four-point amplitude
$A_4(1_\gamma^{-1},2_\gamma^{+1},3_\gamma^{-1},4_\gamma^{+1})$, then
the BCFW approach is feasible. However, if an amplitude contains no
graviton and the naive analysis of Feynman diagrams cannot guarantee that it
will vanish at $z\to\infty$, we don't know whether it can be computed
by the BCFW approach. We will see such an example, namely
$A_6(1_\gamma^{-1},2_\gamma^{+1},3_\gamma^{-1},4_\gamma^{+1},5_\gamma^{-1},6_\gamma^{+1})$.
Secondly, amplitudes of this theory do not have color-order. Thus,
the only difference between external particles which have the same
helicity is their momenta. It means, amplitudes are invariant when
exchanging labels of particles with the same helicity. This symmetry
is useful for calculating and checking results.

The structure of three-point amplitudes indicates that an amplitude of this theory must contain even number of photons, and the sum of helicities of photons is zero. Thus, if we focus on amplitudes containing photons, there are two types of four-point amplitudes, two types of five-point amplitudes, three types of six-point amplitudes, and so on. We will calculate all four-point amplitudes, all five-point amplitudes, and one type of six-point amplitudes, namely $A_6(1_\gamma^{-1},2_\gamma^{+1},3_\gamma^{-1},4_\gamma^{+1},5_\gamma^{-1},6_\gamma^{+1})$. For four-point and five-point amplitudes, we will present more tricks to fix formulae of factorization limits by considering the consistency between different channels. On the other hand, for the six-photon amplitude, we will discuss how to handle one situation: Among all possible deformations, we know the boundary term will appear for some deformations, and don't know whether the boundary term will vanish for other deformations.

\subsection{The four-point amplitude $A_4(1_\gamma^{-1},2_\gamma^{+1},3_g^{-2},4_g^{+2})$}

First let's consider $A_4(1_\gamma^{-1},2_\gamma^{+1},3_g^{-2},4_g^{+2})$. We start from the factorization limit
of $s_{12}\to 0$, and there are two types of solutions
\bea
I_1:~& &\W\la_2=\alpha\W\la_1,~~\la_3=\beta\la_4,~~P_{12}=\gamma\la_4\W\la_1,\nn
& &\alpha={\Spaa{4|1}\over\Spaa{2|4}},~~\beta={\Spbb{1|4}\over\Spbb{3|1}},~~\gamma={\Spaa{1|2}\over\Spaa{4|2}}={\Spbb{3|4}\over\Spbb{1|3}},\nn
I_2:~& &\la_2=\alpha\la_1,~~\W\la_3=\beta\W\la_4,~~P_{12}=\gamma\la_1\W\la_4,\nn
& &\alpha={\Spbb{4|1}\over\Spbb{2|4}},~~\beta={\Spaa{1|4}\over\Spaa{3|1}},~~\gamma={\Spbb{1|2}\over\Spbb{4|2}}={\Spaa{3|4}\over\Spaa{1|3}}.~~\label{4-point-solu}
\eea
For solution $I_1$, we have
\bea
{\lim_{\Spbb{1|2}\to
0}A_3(1_\gamma^{-1},2_\gamma^{+1},-P_g^{-2})A_3(P_g^{+2},3_g^{-2},4_g^{+2})}&=&{\Spaa{1|3}^2\Spaa{2|3}^2\Spbb{4|2}^4\over
s_{13}^2}. \eea
and for solution $I_2$,
\bea
{\lim_{\Spaa{12}\to
0}A_3(1_\gamma^{-1},2_\gamma^{+1},-P_g^{+2})A_3(P_g^{-2},3_g^{-2},4_g^{+2})
}&=&{\Spaa{1|3}^2\Spaa{2|3}^2\Spbb{4|2}^4\over s_{14}^2}. \eea
Now we want to seek a formula $A_{\Spaa{1|2}\Spbb{1|2}}$ which satisfies
factorization limits for both $\Spaa{1|2}\to 0$ and $\Spbb{1|2}\to 0$. It
can be done by rewriting two factorization limits in the
correct expressions. This technique will be used frequently in
latter examples. An useful observation is $s_{13}=-s_{14}$ when
$s_{12}\to 0$, since $s_{12}+s_{13}+s_{14}=0$ for massless
particles. Using this relation, we have
\bea
\lim_{\Spaa{1|2}\to 0}A_LA_R=\lim_{\Spbb{1|2}\to 0}A_LA_R=-{\Spaa{1|3}^2\Spaa{2|3}^2\Spbb{4|2}^4\over s_{13}s_{14}}.
\eea
Thus we can write
\bea
A_{\Spaa{1|2}\Spbb{1|2}}={\Spaa{1|3}^2\Spaa{2|3}^2\Spbb{4|2}^4\over s_{12}s_{13}s_{14}}.
\eea
One can check that $A_{\Spaa{1|2}\Spbb{1|2}}$ has correct factorization limits for remaining channels, therefore it is the correct result of $A_4(1_\gamma^{-1},2_\gamma^{+1},3_g^{-2},4_g^{+2})$. It is the same as the one in \cite{Benincasa:2011kn}.

\subsection{The four-point amplitude $A_4(1_\gamma^{-1},2_\gamma^{+1},3_\gamma^{-1},4_\gamma^{+1})$}

The second amplitude is
$A_4(1_\gamma^{-1},2_\gamma^{+1},3_\gamma^{-1},4_\gamma^{+1})$.
Notice that the naive power counting of Feynman diagrams shows
$A(z\to \infty)\to {1\over z}$ under the deformation of two photons
of same helicity, thus the BCFW approach is feasible although this
case cannot be covered by conclusions in
\cite{ArkaniHamed:2008yf} and \cite{Cheung:2008dn}. Here we try to
find the amplitude using our approach. We will start from considering
the limit $s_{12}\to 0$, where two types of solutions are the same as in
the previous example and we find
\bea A_{\Spbb{1|2}}&=&-{\Spaa{1|3}^2\Spbb{2|4}^2\over s_{12}},~~~~
A_{\Spaa{1|2}}=-{\Spaa{1|3}^2\Spbb{2|4}^2\over s_{12}}. \eea
So we can write
$A_{\Spaa{1|2}\Spbb{1|2}}=A_{\Spaa{1|2}}=A_{\Spbb{1|2}}$ as our starting
expression.

Since $A_{\Spaa{1|2}\Spbb{1|2}}$ does not contain the pole $s_{14}$,
we need to add a new term to get the full answer. The limit  $s_{14}\to
0$ gives
\bea
A_{\Spaa{1|4}\Spbb{1|4}}=-{\Spaa{1|3}^2\Spbb{2|4}^2\over s_{14}}.
\eea
Finally we sum $A_{\Spaa{1|2}\Spbb{1|2}}$ and $A_{\Spaa{1|4}\Spbb{1|4}}$ to get
\bea
A_{\Spaa{1|2}\Spaa{1|4}\Spbb{1|2}\Spbb{1|4}}=A_{\Spaa{1|2}\Spbb{1|2}}+A_{\Spaa{1|4}\Spbb{1|4}}={s_{13}\Spaa{1|3}^2\Spbb{2|4}^2\over s_{12}s_{14}},~~~~\label{PG42F}
\eea
which has correct factorization limits for $s_{12}\to 0$ and $s_{14}\to 0$. It can be verified that $A_{\Spaa{1|2}\Spaa{1|4}\Spbb{1|2}\Spbb{1|4}}$ also has correct factorization limits for remain channels. Thus $A_{\Spaa{1|2}\Spaa{1|4}\Spbb{1|2}\Spbb{1|4}}$ is the correct result for the amplitude $A_4(1_\gamma^{-1},2_\gamma^{+1},3_\gamma^{-1},4_\gamma^{+1})$. It is the same as the one in \cite{Benincasa:2011kn}.

\subsection{The five-point amplitude $A_5(1_\gamma^{-1},2_\gamma^{+1},3_g^{-2},4_g^{+2},5_g^{-2})$}

Now we turn to the five-point amplitude $A_5(1_\gamma^{-1},2_\gamma^{+1},3_g^{-2},4_g^{+2},5_g^{-2})$. Brief analysis of sub-amplitudes shows that this amplitude only contains poles of the form $\Spbb{\bullet|\bullet}$, therefore we need not to consider any channel of the form $\Spaa{\bullet|\bullet}$. We start with the result of the BCFW approach through a deformation which yields the non-zero boundary term. If we deform $1^{-1}$ and $2^{+1}$, the conclusion in \cite{ArkaniHamed:2008yf} indicates that the boundary contribution
will appear. Under the deformation
\bea
\la_2\to \la_2-z\la_1,~~\W\la_1\to \W\la_1+z\W\la_2,
\eea
the BCFW approach gives
\bea
A_{\Spbb{1|3}\Spbb{1|5}}={\Spaa{1|3}\Spaa{4|5}\Spbb{3|4}\Spbb{2|4}^5\over\Spbb{1|3}\Spbb{2|5}\Spbb{3|5}\Spbb{4|5}\Spbb{2|3}^2}
+{\Spaa{1|5}\Spaa{3|4}\Spbb{4|5}\Spbb{2|4}^5\over\Spbb{1|5}\Spbb{2|3}\Spbb{3|4}\Spbb{3|5}\Spbb{2|5}^2},~~\label{A1315}
\eea
which has correct factorization limits for poles $\Spbb{1|3}$ and
$\Spbb{1|5}$. The expression \eref{A1315} is just one representation in the
category and we can deform it to another while keeping correct
factorization limits for poles $\Spbb{1|3}$ and $\Spbb{1|5}$. The
reason to choose another representation is that
\bea
\lim_{\Spbb{2|5}\to 0}s_{25}A_{\Spbb{1|3}\Spbb{1|5}}=\infty,~~\lim_{\Spbb{2|3}\to 0}s_{23}A_{\Spbb{1|3}\Spbb{1|5}}=\infty.
\eea
To remove double poles, we can use following transformations
\bea
\lim_{\Spbb{1|3}\to 0}s_{13}A_{\Spbb{1|3}\Spbb{1|5}}=-{\Spaa{1|3}^2\Spaa{4|5}\Spbb{3|4}\Spbb{2|4}^5\over\Spbb{2|5}\Spbb{3|5}\Spbb{4|5}\Spbb{2|3}^2}
={\Spaa{1|3}^2\Spaa{2|5}\Spbb{3|2}\Spbb{2|4}^5\over\Spbb{2|5}\Spbb{3|5}\Spbb{4|5}\Spbb{2|3}^2}
=-{\Spaa{1|3}^2\Spaa{2|5}\Spbb{2|4}^5\over\Spbb{2|3}\Spbb{2|5}\Spbb{3|5}\Spbb{4|5}},
\eea
and
\bea
\lim_{\Spbb{1|5}\to 0}s_{15}A_{\Spbb{1|3}\Spbb{1|5}}=
-{\Spaa{1|5}^2\Spaa{3|4}\Spbb{4|5}\Spbb{2|4}^5\over\Spbb{2|3}\Spbb{3|4}\Spbb{3|5}\Spbb{2|5}^2}
={\Spaa{1|5}^2\Spaa{3|2}\Spbb{2|5}\Spbb{2|4}^5\over\Spbb{2|3}\Spbb{3|4}\Spbb{3|5}\Spbb{2|5}^2}
=-{\Spaa{1|5}^2\Spaa{2|3}\Spbb{2|4}^5\over\Spbb{2|3}\Spbb{2|5}\Spbb{3|4}\Spbb{3|5}}.
\eea
Then we have
\bea
A'_{\Spbb{1|3}\Spbb{1|5}}={\Spaa{1|3}\Spaa{2|5}\Spbb{2|4}^5\over\Spbb{1|3}\Spbb{2|3}\Spbb{2|5}\Spbb{3|5}\Spbb{4|5}}
+{\Spaa{1|5}\Spaa{2|3}\Spbb{2|4}^5\over\Spbb{1|5}\Spbb{2|3}\Spbb{2|5}\Spbb{3|4}\Spbb{3|5}},
\eea
which will be our starting expression for later calculations.

Since $A_{\Spbb{1|3}\Spbb{1|5}}$ does not contain the pole $\Spbb{1|2}$, we must add a term to give the correct factorization limit
for $\Spbb{1|2}\to 0$. Thus consider $\Spbb{1|2}\to 0$, and we get
\bea
\lim_{\Spbb{1|2}\to 0}A_LA_R&=&-{\Spaa{1|2}^2\Spaa{3|5}\Spbb{1|4}^2\Spbb{2|4}^4\over\Spbb{1|3}\Spbb{1|5}\Spbb{3|4}\Spbb{3|5}\Spbb{4|5}}
=-{\Spaa{1|2}^2\Spaa{3|5}\Spbb{2|4}^6\over\Spbb{2|3}\Spbb{2|5}\Spbb{3|4}\Spbb{3|5}\Spbb{4|5}},\nn
A_{\Spbb{1|2}}&=&-{\lim_{\Spbb{1|2}\to 0}A_LA_R\over s_{12}}=-{\Spaa{1|2}\Spaa{3|5}\Spbb{2|4}^6\over\Spbb{1|2}\Spbb{2|3}\Spbb{2|5}\Spbb{3|4}\Spbb{3|5}\Spbb{4|5}}.
\eea
The purpose of the last step in the first line  is to deform
properly under the limit, so that poles $\Spbb{1|3}$ and $\Spbb{1|5}$ in
the denominator can be removed to keep factorization limits for
$\Spbb{1|3}\to 0$ and $\Spbb{1|5}\to 0$ when $
A'_{\Spbb{1|3}\Spbb{1|5}}$ is added. Now we reach
\bea
A_{\Spbb{1|2}\Spbb{1|3}\Spbb{1|5}}={\Spaa{1|3}\Spaa{2|5}\Spbb{2|4}^5\over\Spbb{1|3}\Spbb{2|3}\Spbb{2|5}\Spbb{3|5}\Spbb{4|5}}
+{\Spaa{1|5}\Spaa{2|3}\Spbb{2|4}^5\over\Spbb{1|5}\Spbb{2|3}\Spbb{2|5}\Spbb{3|4}\Spbb{3|5}}
-{\Spaa{1|2}\Spaa{3|5}\Spbb{2|4}^6\over\Spbb{1|2}\Spbb{2|3}\Spbb{2|5}\Spbb{3|4}\Spbb{3|5}\Spbb{4|5}},
\eea
which has correct factorization limits for poles $\Spbb{1|3}$, $\Spbb{1|5}$, $\Spbb{1|2}$.
One can observe that it is invariant when exchanging $3$ and $5$. It can be checked that it gives correct factorization limits
for remain channels, therefore
\bea
A_5(1_\gamma^{-1},2_\gamma^{+1},3_g^{-2},4_g^{+2},5_g^{-2})=A_{\Spbb{1|2}\Spbb{1|3}\Spbb{1|5}}.~~~~\label{PG51F}
\eea

The calculation above starts from a deformation which yields non-zero boundary contribution.
In order to verify \eref{PG51F}, we can calculate the amplitude by the BCFW approach under a correct deformation, and compare the result with \eref{PG51F}.
Let us choose the deformation as
\bea
\W\la_3\to \W\la_3+z\W\la_5,~~\la_5\to \la_5-z\la_3.
\eea
Then the BCFW approach gives
\bea
A_5(1_\gamma^{-1},2_\gamma^{+1},3_g^{-2},4_g^{+2},5_g^{-2})={\Spbb{1|4}\Spbb{2|4}^5\over\Spbb{1|2}\Spbb{3|5}^2}
\Big({\Spaa{1|3}\Spaa{2|4}\over\Spbb{1|3}\Spbb{2|5}\Spbb{4|5}}
+{\Spaa{3|2}\Spaa{1|4}\over\Spbb{1|5}\Spbb{2|3}\Spbb{4|5}}+{\Spaa{1|2}\Spaa{3|4}\over\Spbb{1|5}\Spbb{2|5}\Spbb{3|4}}\Big).~~~~\label{PG51B}
\eea
One can verify that it is equal to \eref{PG51F} although their expressions
look totally different.

\subsection{The five-point amplitude $A_5(1_\gamma^{-1},2_\gamma^{+1},3_\gamma^{-1},4_\gamma^{+1},5_g^{-2})$}

The next example is the five-point amplitude $A_5(1_\gamma^{-1},2_\gamma^{+1},3_\gamma^{-1},4_\gamma^{+1},5_g^{-2})$. As in the previous case, a brief analysis of sub-amplitudes shows all poles are of the form $\Spbb{\bullet|\bullet}$. We again start from the result
given by the BCFW approach under a wrong deformation which contains the non-zero boundary term. Such a deformation can be chosen as
\bea
\W\la_1\to \W\la_1-z\W\la_2,~~\la_2\to\la_2+z\la_1,
\eea
then the BCFW approach gives
\bea
A_{\Spbb{1|4}\Spbb{1|5}}
=-{\Spaa{1|4}\Spaa{3|5}\Spbb{3|4}\Spbb{2|4}^4\over\Spbb{1|4}\Spbb{2|3}\Spbb{2|5}\Spbb{3|5}\Spbb{4|5}}
+{\Spaa{1|5}\Spaa{3|4}\Spbb{3|5}\Spbb{2|4}^5\over\Spbb{1|5}\Spbb{2|3}\Spbb{3|4}\Spbb{4|5}\Spbb{2|5}^2},
\eea
which produces correct factorization limits for  poles
$\Spbb{1|4}$ and $\Spbb{1|5}$. Again the presence of  the double
pole $\Spbb{2|5}$ suggests us to deform it. To do so,  we can use
$\lim_{\Spbb{1|5}\to 0}\Spaa{3|4}\Spbb{3|5}=-\Spaa{2|4}\Spbb{2|5}$
to rewrite the second term as
\bea
{\Spaa{1|5}\Spaa{3|4}\Spbb{3|5}\Spbb{2|4}^5\over\Spbb{1|5}\Spbb{2|3}\Spbb{3|4}\Spbb{4|5}\Spbb{2|5}^2}\to -{\Spaa{1|5}\Spaa{2|4}\Spbb{2|4}^5\over\Spbb{1|5}\Spbb{2|3}\Spbb{3|4}\Spbb{4|5}\Spbb{2|5}}.
\eea
Thus we get our starting expression
\bea
A'_{\Spbb{1|4}\Spbb{1|5}}=A_{\Spbb{1|4}}+A_{\Spbb{1|5}},
\eea
with
\bea
A_{\Spbb{1|4}}&=&-{\Spaa{1|4}\Spaa{3|5}\Spbb{3|4}\Spbb{2|4}^4\over\Spbb{1|4}\Spbb{2|3}
\Spbb{2|5}\Spbb{3|5}\Spbb{4|5}},~~~
A_{\Spbb{1|5}}=-{\Spaa{1|5}\Spaa{2|4}\Spbb{2|4}^5\over\Spbb{1|5}\Spbb{2|3}\Spbb{3|4}\Spbb{4|5}\Spbb{2|5}}.
~~~\label{A14}\eea
Then we try to include a term to produce the correct factorization
limit for the pole $\Spbb{1|2}$, which is not contained in
$A'_{\Spbb{1|4}\Spbb{1|5}}$. The symmetry of exchanging labels
implies $A_{\Spbb{1|2}}$ can be obtained by exchanging $2$ and $4$
in $A_{\Spbb{1|4}}$, thus we reach
\bea
A_{\Spbb{1|2}\Spbb{1|4}\Spbb{1|5}}&=&A_{\Spbb{1|2}}+A'_{\Spbb{1|4}\Spbb{1|5}}\nn
&=&-{\Spaa{1|2}\Spaa{3|5}\Spbb{2|3}\Spbb{2|4}^4\over\Spbb{1|2}\Spbb{3|4}\Spbb{2|5}\Spbb{3|5}\Spbb{4|5}}
-{\Spaa{1|4}\Spaa{3|5}\Spbb{3|4}\Spbb{2|4}^4\over\Spbb{1|4}\Spbb{2|3}\Spbb{2|5}\Spbb{3|5}\Spbb{4|5}}
-{\Spaa{1|5}\Spaa{2|4}\Spbb{2|4}^5\over\Spbb{1|5}\Spbb{2|3}\Spbb{3|4}\Spbb{4|5}\Spbb{2|5}}.~~\label{A1234}
\eea
Now we consider the factorization limit for the pole $\Spbb{3|4}$.
$A_{\Spbb{3|4}}$ can be obtained by exchanging $1$ and $3$ in
$A_{\Spbb{1|4}}$:
\bea
A_{\Spbb{3|4}}={\Spaa{1|5}\Spaa{3|4}\Spbb{1|4}\Spbb{2|4}^4\over\Spbb{3|4}
\Spbb{1|2}\Spbb{2|5}\Spbb{1|5}\Spbb{4|5}},~~~\label{A34} \eea
thus using \eref{frame-A13} one can construct
\bea
A'_{\Spbb{3|4}}&=&A_{\Spbb{1|2}\Spbb{1|4}\Spbb{1|5}}+(A_{\Spbb{3|4}}-{\lim_{\Spbb{3|4}\to
0}s_{34}A_{\Spbb{1|2}\Spbb{1|4}\Spbb{1|5}}\over s_{34}})\nn
&=&A_{\Spbb{3|4}}+A_{\Spbb{1|4}}.~~~~\label{A34p} \eea
$A'_{\Spbb{3|4}}$ gives correct factorization limits for
$\Spbb{3|4}\to0$ and  $\Spbb{1|4}\to 0$, but no longer has correct
factorization limits for $\Spbb{1|2}\to 0$ and $\Spbb{1|5}\to 0$.
This is a phenomenon one will encounter if the representative expression
is not properly chosen as mentioned in section 2. To avoid
this, one needs to find proper representative expressions for
$A_{\Spbb{3|4}}$ in \eref{A34} and $A_{\Spbb{1|4}}$ in \eref{A14},
such that at each iterative step, the correct factorization limit
is satisfied not only for the new pole, but also for other poles in
previous steps.

Now we try to deform $A_{\Spbb{3|4}}$ in \eref{A34}. Noting that
 $A_{\Spbb{3|4}}$ contains both poles
$\Spbb{1|2}$ and $\Spbb{3|4}$, we try to transform it so that it gives
the correct factorization limit for the pole $\Spbb{1|2}$ automatically.
Using
\bea
\lim_{\Spbb{3|4}\to 0}s_{34}A_{\Spbb{3|4}}&=&-{\Spaa{1|5}\Spaa{3|4}^2\Spbb{1|4}\Spbb{2|4}^4\over\Spbb{1|2}\Spbb{2|5}\Spbb{1|5}\Spbb{4|5}}
={\Spaa{2|5}\Spaa{3|4}^2\Spbb{3|5}\Spbb{2|4}^5\over\Spbb{1|2}\Spbb{2|5}\Spbb{1|5}\Spbb{4|5}\Spbb{3|5}}
=-{\Spaa{1|2}\Spaa{3|4}^2\Spbb{1|3}\Spbb{2|4}^5\over\Spbb{1|2}\Spbb{1|5}\Spbb{2|5}\Spbb{3|5}\Spbb{4|5}},
\eea
and
\bea
\lim_{\Spbb{1|2}\to 0}s_{12}A_{\Spbb{1|2}}&=&{\Spaa{1|2}^2\Spaa{3|5}\Spbb{2|3}\Spbb{2|4}^4\over\Spbb{3|4}\Spbb{2|5}\Spbb{3|5}\Spbb{4|5}}
=-{\Spaa{1|2}^2\Spaa{4|5}\Spbb{1|5}\Spbb{2|4}^5\over\Spbb{3|4}\Spbb{2|5}\Spbb{3|5}\Spbb{4|5}\Spbb{1|5}}
=-{\Spaa{1|2}^2\Spaa{3|4}\Spbb{1|3}\Spbb{2|4}^5\over\Spbb{1|5}\Spbb{2|5}\Spbb{3|4}\Spbb{3|5}\Spbb{4|5}},
\eea
we obtain
\bea
A_{\Spbb{1|2}\Spbb{3|4}}=A_{\Spbb{1|2}}=A_{\Spbb{3|4}}={\Spaa{1|2}\Spaa{3|4}\Spbb{1|3}
\Spbb{2|4}^5\over\Spbb{1|2}\Spbb{1|5}\Spbb{2|5}\Spbb{3|4}\Spbb{3|5}\Spbb{4|5}},~~~\label{new-A34}
\eea
where $A_{\Spbb{1|2}}$ in \eref{A1234} has been deformed as well.
Using the deformed $A_{\Spbb{3|4}}$ in \eref{new-A34} from \eref{A1234} to
\eref{A34p}, we find that the new $A'_{\Spbb{3|4}}$ gives correct
factorization limits for poles $\Spbb{3|4}$,  $\Spbb{1|4}$, and
$\Spbb{1|2}$, but not for the pole $\Spbb{1|5}$. To fix this problem, we
need to deform $A_{\Spbb{1|4}}$ or  $A_{\Spbb{1|5}}$ in \eref{A14}.

Now noticing the symmetry  $1\leftrightarrow 3$ or $2\leftrightarrow
4$, we can construct  $A_{\Spbb{1|4}\Spbb{2|3}}$ from
$A_{\Spbb{1|2}\Spbb{3|4}}$ in \eref{new-A34} by exchanging $2$ and
$4$ as
\bea
A_{\Spbb{1|4}\Spbb{2|3}}=A_{\Spbb{1|4}}=A_{\Spbb{2|3}}=-{\Spaa{1|4}\Spaa{2|3}\Spbb{1|3}\Spbb{2|4}^5\over\Spbb{1|4}
\Spbb{1|5}\Spbb{2|3}\Spbb{2|5}\Spbb{3|5}\Spbb{4|5}}.~~~\label{new-A14}
\eea
Putting the new expressions in \eref{new-A34} and \eref{new-A14}
into \eref{A34p},  we reach
\bea
A_{\Spbb{1|2}\Spbb{1|4}\Spbb{2|3}\Spbb{3|4}}=A_{\Spbb{1|2}\Spbb{3|4}}+A_{\Spbb{1|4}\Spbb{2|3}},
\eea
which gives correct factorization limits for $\Spbb{1|2}\to 0$, $\Spbb{1|4}\to 0$, $\Spbb{2|3}\to 0$,
 $\Spbb{3|4}\to 0$ as well as $\Spbb{1|5}\to 0$.
To check it, notice that

\bea
\lim_{\Spbb{1|5}\to 0}s_{15}A_{\Spbb{1|2}\Spbb{1|4}\Spbb{2|3}\Spbb{3|4}}&=&{\Spaa{1|5}\Spaa{1|4}\Spaa{2|3}\Spbb{1|3}\Spbb{2|4}^5
\over\Spbb{1|4}\Spbb{2|3}\Spbb{2|5}\Spbb{3|5}\Spbb{4|5}}
-{\Spaa{1|5}\Spaa{1|2}\Spaa{3|4}\Spbb{1|3}\Spbb{2|4}^5\over\Spbb{1|2}\Spbb{2|5}\Spbb{3|4}\Spbb{3|5}\Spbb{4|5}}\nn
&=&{\Spaa{1|5}\Spaa{2|4}\Spbb{2|4}^5\over\Spbb{2|5}\Spbb{3|5}\Spbb{4|5}}\Big({\Spaa{1|2}\over\Spbb{3|4}}-{\Spaa{1|4}\over\Spbb{2|3}}\Big)\nn
&=&{\Spaa{1|5}\Spaa{2|4}\Spbb{2|4}^5\Spab{1|2+4|3}\over\Spbb{2|3}\Spbb{2|5}\Spbb{3|4}\Spbb{3|5}\Spbb{4|5}}\nn
&=&{\Spaa{1|5}\Spaa{2|4}\Spbb{2|4}^5\Spaa{1|5}\Spbb{3|5}\over\Spbb{2|3}\Spbb{2|5}\Spbb{3|4}\Spbb{3|5}\Spbb{4|5}},
\eea
therefore
\bea
\lim_{\Spbb{1|5}\to 0}s_{15}A_{\Spbb{1|2}\Spbb{1|4}\Spbb{2|3}\Spbb{3|4}}=s_{15}A_{\Spbb{1|5}}=-\lim_{\Spbb{1|5}\to 0}A_LA_R.
\eea
One can verify that
$A_{\Spbb{1|2}\Spbb{1|4}\Spbb{2|3}\Spbb{3|4}}$ also gives correct
factorization limits for remaining channels. Thus we find the correct
amplitude is
\bea
A(1_\gamma^{-1},2_\gamma^{+1},3_\gamma^{-1},4_\gamma^{+1},5_g^{-2})=A_{\Spbb{1|2}\Spbb{1|4}\Spbb{2|3}\Spbb{3|4}}.~~~~\label{PG52F}
\eea
Notice that the formula of $A_{\Spbb{1|2}\Spbb{1|4}\Spbb{2|3}\Spbb{3|4}}$ is manifestly invariant when
exchanging $1\leftrightarrow 3$ and $2\leftrightarrow 4$. Result \eref{PG52F} can be rewritten as
\bea
A(1_\gamma^{-1},2_\gamma^{+1},3_\gamma^{-1},4_\gamma^{+1},5_g^{-2})&=&{\Spbb{1|3}\Spbb{2|4}^5(
\Spaa{1|4}\Spbb{3|4}\Spba{1|4+5|3}-\Spbb{1|4}\Spaa{3|4}\Spab{1|4+5|3})
\over\Spbb{1|2}\Spbb{1|4}\Spbb{1|5}\Spbb{2|3}\Spbb{2|5}\Spbb{3|4}\Spbb{3|5}\Spbb{4|5}}\nn
&=&{\Spbb{1|3}\Spbb{2|4}^5(\Spbb{1|4}\Spbb{3|5}\Spaa{1|5}\Spaa{3|4}-\Spaa{1|4}\Spaa{3|5}\Spbb{1|5}\Spbb{3|4})
\over\Spbb{1|2}\Spbb{1|4}\Spbb{1|5}\Spbb{2|3}\Spbb{2|5}\Spbb{3|4}\Spbb{3|5}\Spbb{4|5}},~~~~\label{PG52B}
\eea
which is the formula in \cite{Benincasa:2011kn}.

As a sidenote, when two deformed particles are two photons of
same helicity, there is no general proof of its boundary
behavior, since this situation cannot be covered by conclusions in
\cite{ArkaniHamed:2008yf, Cheung:2008dn}. Although naive power
counting of Feyman diagrams shows that the large $z$ behavior is $z^0$,
using the result given in \eref{PG52B}, one can observe that under
the deformation of two photons of same helicity, the boundary
term will vanish.

\subsection{The six-point amplitude $A_6(1_\gamma^{-1},2_\gamma^{+1},3_\gamma^{-1},4_\gamma^{+1},5_\gamma^{-1},6_\gamma^{+1})$}

The final example is the six-point  amplitude
$A_6(1_\gamma^-,2_\gamma^+,3_\gamma^-,4_\gamma^+,5_\gamma^-,6_\gamma^+)$.
This case is a little different from those in previous subsections.
For previous cases, we know there exist some deformations that can make
boundary terms vanish, so we do not need our approach in
this paper to find them. However, for
$A_6(1_\gamma^{-1},2_\gamma^{+1},3_\gamma^{-1},4_\gamma^{+1},5_\gamma^{-1},6_\gamma^{+1})$
the situation is different. First, since there is no graviton, results
in \cite{Cheung:2008dn} cannot be applied.
Secondly, we know the boundary term is not zero when deforming two
photons with opposite helicities. Thirdly,  when two deformed
photons have the same helicity, the large $z$ behavior is $z^0$ by
naive power counting of Feynman diagrams. Thus, our approach becomes
one of useful approaches.

Let us consider the deformation:
\bea
\la_1\to \la_1-z\la_3,~~\W\la_3\to \W\la_3+z\W\la_1.
\eea
Under this, the BCFW approach gives residues for the following poles: $\Spbb{1|2}$, $\Spbb{1|4}$, $\Spbb{1|6}$,
$\Spaa{3|2}$, $\Spaa{3|4}$, $\Spaa{3|6}$, $s_{124}$, $s_{125}$, $s_{126}$, $s_{145}$, $s_{146}$, $s_{156}$. A little bit algebra yields
\bea
A&=&{\Spaa{1|2}\Spaa{4|6}\Spab{5|1+3|2}^5(\Spaa{4|5}\Spbb{3|4}\Spbb{5|6}\Spab{6|1+3|2}-\Spaa{5|6}\Spbb{3|6}\Spbb{4|5}\Spab{4|1+3|2})
\over \Spbb{1|2}\Spaa{4|5}\Spaa{5|6}\Spab{4|1+3|2}\Spab{6|1+3|2}\Spab{4|1+2|3}\Spab{5|1+2|3}\Spab{6|1+2|3}s_{123}}+(2\leftrightarrow 4)+(2\leftrightarrow 6)\nn
& &+{\Spaa{1|3}^4\Spbb{2|3}\Spbb{4|6}^5\Spab{2|1+3|5}(\Spaa{1|4}\Spaa{5|6}\Spbb{4|5}\Spab{2|1+3|6}-\Spaa{1|6}\Spaa{4|5}\Spbb{5|6}\Spab{2|1+3|4})
\over \Spaa{2|3}\Spbb{4|5}\Spbb{5|6}\Spab{2|1+3|4}\Spab{2|1+3|6}\Spab{1|2+3|4}\Spab{1|2+3|5}\Spab{1|2+3|6}s_{123}}\nn
& &+(2\leftrightarrow 4)+(2\leftrightarrow 6)\nn
& &+{s_{24}\Spaa{5|6}\Spbb{3|5}\Spab{1|2+4|6}^5\over\Spaa{1|2}\Spaa{1|4}\Spbb{3|6}\Spbb{5|6}\Spab{1|2+4|5}\Spab{2|1+4|3}\Spab{4|1+2|3}s_{124}}
+(4\leftrightarrow 6)+(2\leftrightarrow 6)\nn
& &+{s_{46}\Spaa{1|5}^5\Spbb{2|5}\Spbb{4|6}^4\Spab{2|1+5|3}\over \Spaa{1|2}\Spaa{2|5}\Spbb{3|4}\Spbb{3|6}
\Spab{1|2+5|4}\Spab{1|2+5|6}\Spab{5|1+2|3}s_{125}}+(2\leftrightarrow 4)+(2\leftrightarrow 6).~~~~
\label{PG61B}
\eea
Four explicit terms above are residues for poles
$\Spbb{1|2}$, $\Spaa{2|3}$, $s_{124}$ and $s_{125}$. Other terms can be
obtained by exchanging labels. Result \eref{PG61B} will be our
starting expression.

Now we need to include factorization limits for poles not detected by the chosen deformation.
It turns out that result \eref{PG61B} is the correct amplitude that we are
seeking for, since it gives correct factorization limits for all
physical channels. The symmetry of exchanging labels makes the verification very
easy. The amplitude has the $S_3\bigotimes S_3$ symmetry:
$1\leftrightarrow 3$, $3\leftrightarrow 5$, $5\leftrightarrow 1$ and
$2\leftrightarrow 4$, $4\leftrightarrow 6$, $6\leftrightarrow 2$.
Its factorization limits will be restricted by this symmetry, for
example, $\lim_{\Spbb{2|3}\to 0}A_LA_R$ can be obtained by exchanging
$1$ and $3$ in $\lim_{\Spbb{1|2}\to 0}A_LA_R$. \eref{PG61B} already
has correct factorization limits for poles detected by the BCFW
approach. If it is invariant when some labels are exchanged, more correct
factorization limits will be satisfied. For instance, \eref{PG61B}
provides the correct factorization limit for $\Spbb{1|2}\to 0$. If we
perform the exchange $1\leftrightarrow 3$ for \eref{PG61B}, the
new expression gives the correct factorization limit for
$\Spbb{2|3}\to 0$, since $\lim_{\Spbb{2|3}\to 0}A_LA_R$ can be
obtained from $\lim_{\Spbb{1|2}\to 0}A_LA_R$ by this exchange.
Thus, if \eref{PG61B} has the symmetry $1\leftrightarrow 3$, the
correct factorization limit for $\Spbb{2|3}\to 0$ is also satisfied.
Consequently, if \eref{PG61B} has the $S_3\bigotimes S_3$ symmetry
as the correct amplitude, all factorization limits will be given. In
\eref{PG61B}, the symmetry $\{2\leftrightarrow
4,~4\leftrightarrow 6,~6\leftrightarrow 2\}$ is manifest. On the
other hand, we have checked the symmetry $\{1\leftrightarrow
3,~3\leftrightarrow 5,~5\leftrightarrow 1\}$ numerically. Further more, we have verified that all spurious poles are canceled, as we did in section
4.2. Therefore
\eref{PG61B} is the correct
$A_6(1_\gamma^{-1},2_\gamma^{+1},3_\gamma^{-1},4_\gamma^{+1},5_\gamma^{-1},6_\gamma^{+1})$.

\section{Example 4: Yukawa theory (Elimination of spurious poles)}

In this section we will show how to treat the case that an expression has correct factorization limits for all
physical poles, but contains spurious poles. We have not encounter such phenomenon in previous examples.
However, the situation will appear when considering the color ordered amplitude of fermions coupling to scalars by Yukawa coupling.

In this theory, three-point amplitudes are given by
\bea
A_3(1^+,2,3^+)=\Spbb{1|3},~~A_3(1^-,2,3^-)=\Spaa{1|3},~~A_3(1^+,2,3^-)=A_3(1^-,2,3^+)=0,
\eea
where the coupling constant has been neglected.
Let us consider the simplest case, the four point amplitude $A_4(1^-,2,3,4^+)$. This amplitude corresponding to only one Feynman diagram and the
result can be obtained by Feynman rules as $^{[7]}$\footnotetext[7]{Since the propagator ${i\not P\over P^2}$ depend on the direction of $P$,
we assume that the fermion line is from $4$ to $1$ and all external momenta are coming.}
\bea
A_4(1^-,2,3,4^+)={\Spab{1|P_{34}|4}\over s_{34}}={\Spaa{1|3}\over\Spaa{4|3}}=-{\Spbb{2|4}\over\Spbb{2|1}}.
\eea
Now we use our approach to reproduce this result. Physical poles come from $s_{12}\to 0$, and only the solution $\W\la_2\propto\W\la_1$
provides non-vanishing sub-amplitudes, thus there is only one physical pole $\Spbb{1|2}$. Using solution $I_1$ in \eref{4-point-solu},
we get the factorization limit for this pole as
\bea
A_{\Spbb{1|2}}=-{\lim_{\Spbb{1|2}\to 0}A_LA_R\over s_{12}}={s_{14}\over \Spaa{2|4}\Spbb{1|2}}.
\eea
The expression $A_{\Spbb{1|2}}$ which has correct factorization limit for the physical pole but also contains a spurious pole
$\Spaa{2|4}$. To eliminate it, let us use the approach discussed in section 2, \textit{i.e.}, choose a correct expression for
${\lim_{\Spaa{2|4}\to 0}\Spaa{2|4}A_{\Spaa{1|2}}\over \Spaa{2|4}}$ so that it does not contain any physical pole. This procedure can be done as
\bea
\lim_{\Spaa{2|4}\to 0}\Spaa{2|4}A_{\Spaa{1|2}}=\lim_{\Spaa{2|4}\to 0}{s_{14}\over \Spbb{1|2}}={-s_{12}\over \Spbb{1|2}}=\Spaa{1|2}.
\eea
Then we can construct
\bea
A'_{\Spbb{1|2}}=A_{\Spbb{1|2}}-{\lim_{\Spaa{2|4}\to 0}\Spaa{2|4}A_{\Spaa{1|2}}\over \Spaa{2|4}}={-s_{13}\over\Spaa{2|4}\Spbb{1|2}}
={\Spbb{2|4}\over\Spbb{1|2}},
\eea
which is the correct answer.

However, above approach for removing spurious poles will be difficulty to perform if the amplitude contains many physical poles. We have not find a
more efficient approach.

\section{Conclusion \label{6}}

In this paper, we have proposed an approach to calculate tree
amplitudes without polynomial terms through their factorization limits. We seek for a quantity
that has correct factorization limits for all physical poles and does not contain other poles
iteratively. Starting from an initial function which has correct
factorization limits for some poles, we adjust our expression to
include factorization limits for new channels at each
iterative step, while keeping correct factorization  limits of
previous poles. Proceeding by this, the proper choice of an
expression in the equivalent category under corresponding limits is
required. We have shown how to make such choice in various examples.
Because at each step, at least one new pole is included into the
set of channels having correct factorization limits, this algorithm
will stop at finite steps. After obtaining an expression which has correct factorization limits
for all physical poles, we need to eliminate possible spurious poles. Then we get the desired result. This approach can be applied to all
circumstances no matter whether the boundary contribution vanishes. However, this approach cannot determine
polynomial terms since it can only detect the pole part. If the amplitude admits polynomials which satisfy correct mass dimension and helicities, the full amplitude cannot be determined.

To demonstrate, we have applied our approach to calculate amplitudes
of $\phi^4$ theory, pure gauge theory, Einstein-Maxwell theory, Yukawa theory. Correct results of these examples are
obtained, although their calculations are somewhat complicated. In these
examples, one can see that no information about boundary terms is
required when using this approach.

In principle, one can split an amplitude into more than two
sub-amplitudes by imposing  on-shell conditions on more propagators.
However, it will make the computation more complicated. For
example, if we cut the amplitude into three sub amplitudes, we need
to consider factorization limits for all possible combinations of
two channels. The number of such combinations grows extremely faster
than the number of channels. The maximal cut is imposing on-shell
conditions on all propagators, then all sub amplitudes are
lowest-point amplitudes. In such case, all possible Feynman diagrams
need to be considered one by one.

It is interesting to consider whether this approach can be generalized to tree amplitudes
of string theory. The most difficult issue is, in string theory the number of inner states is infinite.
We have not found a way to tackle this difficulty.

In this paper, all examples are within consistent quantum field theories. Another important direction in future is to applied this approach to any theory which is known to be inconsistent. An example is  massless spin-3 fields: here the three point amplitudes are known, but on general grounds no higher point tree level amplitudes should be `constructable', thus our approach encounters the difficulty. Also, coupling gauge or gravity fields to higher spin ($>2$) massive or massless matters would be interesting.

\section*{Acknowledgement}

The authors would like to thank Prof. Bo Feng for his guidance throughout the
project.  This work is supported by Chinese NSF funding under
contracts No.11031005, No.11135006 and No.11125523.

\appendix

\section{Absence of polynomials}

Here we give a brief proof of that amplitudes calculated in this paper do not contain any polynomial term.
The argument is simple: One cannot construct a polynomial that has correct mass dimension of the amplitude
and correct helicities of all external particles.

The mass dimension of a bare amplitude is $D=4-n-\sum D_c$. Here $D_c$ are mass dimensions of coupling constants. When we mention the `bare amplitude', we mean that all coupling constants are stripped off. For amplitudes of $\phi^4$ theory, pure gauge theory and Yukawa theory,
since the coupling constants are dimensionless, we have $D<0$ if the amplitude contains at least five external particles, and $D=0$ if the amplitude
contains four external particles. Thus, the only possible polynomial is a constant which corresponding to four-point amplitudes.
However, a constant cannot provide correct helicities for external particles, unless all external states are scalars. Consequently, for such amplitudes, the only allowed polynomial is just the lowest-point amplitude of $\phi^4$ theory.

For amplitudes of Einstein-Maxwell theory, the coupling constant has mass dimension $-1$, thus for all amplitudes of this theory we
have $D=-2$ for an arbitrary $n$. Then the possible polynomial can take the form $\Spaa{\bullet|\bullet}\Spaa{\bullet|\bullet}$,
$\Spbb{\bullet|\bullet}\Spbb{\bullet|\bullet}$, and $\Spaa{\bullet|\bullet}\Spbb{\bullet|\bullet}$. None of these can provide correct
helicities for all external states since one spinor can only carries helicity $\pm{1\over 2}$, and the amplitudes we have calculated
contain at least four external particles (recall that a photon has helicity $\pm1$, one graviton has helicity $\pm2$, and for spinorial products, $\Spaa{i|i}=0,~\Spbb{i|i}=0$).

Hence, all amplitudes mentioned in this paper do not contain any polynomial term.

\section{Alternative calculation of $A_6(1^-,2^-,3^-,4^+,5^+,6^+)$\label{A}}

In this section, we present the calculation of
$A_6(1^-,2^-,3^-,4^+,5^+,6^+)$ that starts by considering the
factorization limit for the three-particle channel $s_{234}\to 0$.
Unlike the two-particle channel where the on-shell limit is split into the
holomorphic and anti-holomorphic parts, the on-shell limit of
$s_{234}\to 0$ could not be split, therefore $s_{234}$ will appear in the
denominator of the amplitude as a whole. Thus we do not solve the
constraint on kinematic variables, as the calculation in
the $\phi^4$ case. The factorization limit for $s_{234}\to 0$
is given by
\bea
\lim_{s_{234}\to 0}A_4(2^-,3^-,4^+,-P_{234}^+)A_4(P_{234}^-,5^+,6^+,1^-)&=&{\Spaa{2|3}^3\over\Spaa{3|4}\Spaa{4|P_{234}}\Spaa{P_{234}|2}}
{\Spaa{1|P_{234}}^3\over\Spaa{P_{234}|5}\Spaa{5|6}\Spaa{6|1}}.
\eea
To express $\la_{P_{234}}$, we can use $\Spaa{a|P_{234}}= {\Spab{a|P_{234}|b}\over \Spbb{P_{234}|b}}$.
The calculation is following,
\bea
\lim_{s_{234}\to 0}A_4(2^-,3^-,4^+,-P_{234}^+)A_4(P_{234}^-,5^+,6^+,1^-)&=&
{\Spaa{2|3}^3\over\Spaa{3|4}\Spaa{4|P_{234}}\Spaa{P_{234}|2}}{\Spaa{1|P_{234}}^3\over\Spaa{P_{234}|5}\Spaa{5|6}\Spaa{6|1}}
{\Spbb{P_{234}|2}^2\Spbb{P_{234}|4}\over \Spbb{P_{234}|2}^2\Spbb{P_{234}|4}}\nn
&=&{\Spab{1|P_{234}|4}\Spaa{1|P_{234}2|3}^2\over\Spaa{5|6}\Spaa{6|1}\Spaa{3|4}^2\Spbb{2|3}\Spbb{3|4}\Spab{5|P_{234}|2}}\nn
&=&{\Spab{1|2+3|4}\Spaa{1|P_{234}4|3}^2\over\Spaa{5|6}\Spaa{6|1}\Spaa{3|4}^2\Spbb{2|3}\Spbb{3|4}\Spab{5|3+4|2}}\nn
&=&{\Spab{1|2+3|4}^3\over\Spaa{5|6}\Spaa{6|1}\Spbb{2|3}\Spbb{3|4}\Spab{5|3+4|2}}.
\eea
From the second line to the third line, we have used
$\Spaa{1|P_{234}2|3}=\Spaa{1|P_{234}(2+3)|3}=-\Spaa{1|P_{234}4|3}$.
This step is necessary if we try to include correct factorization
limits for poles $\Spaa{5|6}$ and $\Spbb{2|3}$. Thus we obtain
\bea
A_{\Spaa{5|6}\Spbb{2|3}s_{234}}=-{\Spab{1|2+3|4}^3\over\Spaa{5|6}\Spaa{6|1}\Spbb{2|3}\Spbb{3|4}\Spab{5|3+4|2}s_{234}}.~~\label{part1}
\eea
Since $A_{\Spaa{5|6}\Spbb{2|3}s_{234}}$ does not contain the pole $s_{126}$, we should compute the factorization limit for this pole and add it to $A_{\Spaa{5|6}\Spbb{2|3}s_{234}}$.
Similar manipulation gives
\bea
A_{\Spaa{4|5}\Spbb{1|2}s_{126}}=-{\Spab{3|4+5|6}^3\over\Spaa{3|4}\Spaa{4|5}\Spbb{1|2}\Spbb{6|1}\Spab{5|3+4|2}s_{126}}.~~\label{part2}
\eea
Summing \eref{part1} and \eref{part2}, we reach the correct result.


\end{document}